\documentclass[11pt,english]{article}
\usepackage[T1]{fontenc}
\usepackage[latin9]{inputenc}
\usepackage{geometry}
\usepackage{babel}
\usepackage{rotfloat}
\usepackage{amsmath}
\usepackage{amsthm}
\usepackage{graphicx}
\usepackage{setspace}
\usepackage[unicode=true]
 {hyperref}

\makeatletter

\usepackage{MnSymbol}%
\usepackage{wasysym}%

\usepackage{xcolor}

\usepackage{harvard}
\usepackage{booktabs}

\@ifundefined{showcaptionsetup}{}{%
 \PassOptionsToPackage{caption=false}{subfig}}
\usepackage{subfig}
\makeatother

\begin{document}

\title{Resilient Urban Housing Markets:\\
Shocks vs. Fundamentals\thanks{I would like to thank the Johns Hopkins Coronavirus Research Center,
the Inter-university Consortium for Political and Social Research,
Zillow, the OpenStreetMap Foundation, Microsoft Research, for access
to microdata. I would like to thank Ambika Gandhi for her careful
comments on the manuscript.}}

\author{Amine Ouazad\thanks{Associate Professor of Economics, Research Professorship in ``Urban
and Real Estate Economics'', HEC Montreal. 3000, Chemin de la Côte
Sainte Catherine, Montreal H3T 2A7, Canada. amine.ouazad@hec.ca.}}

\date{September 2020}
\maketitle
\begin{abstract}
In the face of a pandemic, urban protests, and an affordability crisis,
is the desirability of dense urban settings at a turning point? Assessing
cities\textquoteright{} long term trends remains challenging. The
first part of this chapter describes the short-run dynamics of the
housing market in 2020. Evidence from prices and price-to-rent ratios
suggests expectations of resilience. Zip-level evidence suggests a
short-run trend towards suburbanization, and some impacts of urban
protests on house prices. The second part of the chapter analyzes
the long-run dynamics of urban growth between 1970 and 2010. It analyzes
what, in such urban growth, is explained by short-run shocks as opposed
to fundamentals such as education, industrial specialization, industrial
diversification, urban segregation, and housing supply elasticity.
This chapter\textquoteright s original results as well as a large
established body of literature suggest that fundamentals are the key
drivers of growth. The chapter illustrates this finding with two case
studies: the New York City housing market after September 11, 2001;
and the San Francisco Bay Area in the aftermath of the 1989 Loma Prieta
earthquake. Both areas rebounded strongly after these shocks, suggesting
the resilience of the urban metropolis.
\end{abstract}
\clearpage{}

\pagebreak{}

\section{Introduction}

Between 55\% (United Nations Population Division) and 85\% (European
Commission) of world population lives in urban areas. Such population
is concentrated on a small share of the world's landmass: between
0.45\% (Liu, He, Zhou, Wu, 2014) and 1.5\% (European Commission) depending
on the estimates. The spatial concentration of location choices can
be explained by agglomeration economies: a key mechanism that enables
the description of the spatial distribution of location choices and
economic activity using the tools of general equilibrium. The basic
mechanisms of agglomeration economies were described as early as in
\possessivecite{marshall1890principles} Principles of Economics,
and are the essential ingredient in spatial models, including \citeasnoun{fujita1996economics}
and \citeasnoun{behrens2015agglomeration}. Agglomeration enables
the sharing of common resources, the matching with potential employers,
buyers, sellers, partners; it also enables learning and social interactions.
Dense urban living makes workers more productive~\cite{puga2010magnitude}.
Agglomeration economics underpin the emergence and growth of cities
\cite{duranton2004micro}. Efforts to estimate the magnitude of agglomeration
economies are described in \citeasnoun{rosenthal2004evidence}, \citeasnoun{melo2009meta}
and \citeasnoun{combes2015empirics}.

Recent events have raised concerns that the benefits of agglomeration
may be declining, affecting the desirability of urban living; perhaps
even triggering an exodus from cities. The high density of urban setting
suggests that, over long periods of time, the benefits of agglomeration
have typically outstripped the costs of living in urban settings.
These include traffic congestion \cite{duranton2011fundamental},
potential health hazards \cite{moore2003global}\footnote{The causal impact of urbanization on health is ambiguous. For instance
\citeasnoun{singh2014widening} displays a life expectancy that is
2.7 year longer in urban areas vs rural areas of the United States.
Urbanization can lead to worse health outcomes in urban slums (Riley,
Ko, Unger, Reis). While statistical correlations also suggest that
urbanization is a necessary condition for growth, there are examples
of urbanization without growth, e.g. in Sub-Saharan Africa and in
South Asia~(Annez and Buckley, Chapter 1 in Spence, Annez, Buckley,
2009; Chauvin, Glaeser, Ma, Tobio, 2017).}, labor poaching~\cite{combes2006labour}. 

At least two shocks have affected urban areas in 2020: the Covid-19
pandemic and urban protests. Anecdotal evidence, statements by public
officials as well as descriptive statistics suggest a positive correlation
between urban density and the number of confirmed Covid-19 cases per
capita.\footnote{Table~\ref{tab:regression_drivers_covid19_infections} presents regressions
suggesting a statistically significant positive correlation between
county population density and confirmed cases per capita.} In addition, urban protests focusing on racial justice have taken
place in 43\% of the 917 metropolitan areas in May 2020.\footnote{This statistic uses geocoded protest location data and Zillow's definition
of metropolitan area boundaries. These data are described in Section~\ref{subsec:Local-Housing-Markets}.} Thus a key question is whether the multiple short-run shocks to urban
housing markets are likely to cause a long-term decline in metropolitan
population growth. Will advances in information technology coupled
with the challenges of living in dense neighborhoods lead to a decline
of urban living, with a population living farther away from the densest
cities? The answer is ultimately an empirical question, that can be
informed by the analysis of (i)~the nature of recent short-run \emph{shocks}
to local housing demand, and (ii)~the importance of short-run shocks
versus long-run \emph{fundamentals} for the growth of metropolitan
areas.

This chapter presents an analysis of the \emph{short-run} shocks to
the housing market in 2020 using Zip-level housing data. As the long-term
prospects of U.S. urban housing markets cannot yet be assessed, the
chapter turns to the past to inform the future. The chapter performs
an analysis of the \emph{long-term} 1970-2010 growth trends of 306
metropolitan areas. It then presents two Zip-level case studies of
the long-term resilience of New York City's housing market after September
11, 2001, and of the long-term resilience of the San Francisco Bay
area after the 1989 Loma Prieta earthquake. 

By combining micro data on Covid-19 infections, geocoded urban protests
with census demographics, house prices, inventories, and rents, the
chapter documents the large magnitude of the series of shocks that
affected U.S. housing markets in 2020: prices, inventories, rents
all experienced large movements. Yet, despite such large shocks, the
dynamic of prices is consistent with the market's \emph{expectations
of resilience}. There is also evidence that, within metropolitan areas,
housing demand is increasing faster in less dense neighborhoods and
in neighborhoods farther away from the center of the metropolitan
area. This is consistent with, at least in the short-run, households'
\emph{adaptation} to changing conditions by demanding housing in locations
farther away from the impact of the short-run shocks.

The second part of the chapter uses longitudinal time series of census
tracts with consistent 2010 boundaries to estimate the impact of fundamentals
and shocks on the population growth in 306 metropolitan areas.\footnote{Recent data includes information on more than 900 metropolitan areas.
The 1970-2010 longitudinal data of the Neighborhood Change Database
allows an analysis of 306 metropolitan areas.} Results suggest no statistically significant impact of shocks such
as hurricanes and urban protests. This may be surprising given the
experience of New Orleans: in 2018 population was 16\% lower than
its pre-hurricane 2005 level. Yet, in other metropolitan areas, billion
dollar events such as hurricane Harvey and hurricane Sandy had no
discernible impact on metropolitan population levels. Beyond differences
in the hydrology and topography of New Orleans, Houston, and New York,
a set of economic differences in fundamentals may explain the divergent
long-term paths in response to short-run shocks. This may also explain
why \citeasnoun{collins2007economic} finds a long-run impact of the
1967 Detroit riots (a short-run shock) on long-run population growth
and on property values. Detroit's population peaked in 1950, 17 years
before the riots. \citeasnoun{glaeser2011triumph} argues that Detroit's
industrial mono-culture may have hindered innovation. Hence the shock
of the riots may have been correlated with or driven by economic fundamentals
such as Detroit's relatively lower industrial diversification and
high level of racial segregation. This chapter's description of the
importance of shocks vs. fundamentals does not establish that shocks
do not affect metropolitan population growth (\citeasnoun{boustan2020effect}
describes impacts on outmigration and income) but rather that fundamentals
may outweigh their impacts.

The chapter also presents a case study of the resilience of local
housing markets in the aftermath of September 11, 2001 in New York.
There is evidence of a short-run reversal of the gradient between
price appreciation and distance to the Central Business District during
the September\textendash December 2001 period. Yet the gradient returns
to its prior, long-term, negative slope whereby price appreciation
is higher in the CBD. There is no impact of the event on house price
appreciation from 2002 onward. Similar findings emerge in this chapter's
second case study, the impact of the 1989 Loma Prieta earthquake on
San Francisco's housing markets. While there are visible population
outflows in the 1990 Census in areas affected by the earthquake, there
is no long term difference in population trends across areas with
different earthquake risks. Large corporate headquarters have sprung
up in those areas at risk of earthquakes. Overall the metro-level
and the neighborhood-level evidence are consistent with the resilience
of urban housing markets, whereby fundamentals drive metropolitan
growth rather than short-run shocks.

This chapter's findings are consistent with prior literature. \citeasnoun{davis2002bones}
documents the evolution of Japanese cities from the Stone Age to the
modern era, with a specific focus on the impacts of World War II bombing
on the growth of Japanese cities. They document a strong recovery
in the years immediately after such an unprecedented shock to city
population. \citeasnoun{brakman2004strategic} documents that this
is also true of German cities strategically bombed during World War
II: the impact on city growth is only temporary. \citeasnoun{davis2002bones}
emphasizes the importance of locational fundamentals. This chapter
emphasizes the importance of fundamentals such as education, industrial
composition, and urban segregation. In other words, it can be argued
that while resources such as coal or proximity to major streams may
have determined the emergence of cities, it is their education levels,
their diverse economic activity, and the opportunity to interact and
learn that is the modern foundation of urban living. 

This chapter's results also suggest that housing supply elasticity
is a positive driver of metropolitan population growth. Limited housing
supply elasticity in some metropolitan areas might be driving recent
population outflows from California to more affordable housing markets
in Texas, Arizona, Nevada, and other states. \citeasnoun{zabel2012migration}
finds that the cost of housing is a driver of labor mobility across
metropolitan areas during the 1990-2006 period. Limited housing supply
elasticity may be hindering recovery after shocks. \citeasnoun{koster2012bombs}
argues that planning policies may have hindered the rebuilding of
bombed areas in Rotterdam after World War II. 

Finally, this chapter's findings are also consistent with prior work
on pandemics and housing markets. \citeasnoun{korevaar2020} finds
only short-run impacts on house prices and rents of the 17th century
plague in Amsterdam of 19th century cholera in Paris. These effects
are short-lived as they do not last more than a year. These results
are also consistent with the Canadian experience of the SARS pandemic.
On April 23, 2003, the World Health Organization issued a travel advisory
for Toronto recommending postponing all but essential travel. There
is however no evidence of impacts of SARS infections on the growth
of Toronto's housing markets (prices and transaction volumes) in 2003
and beyond. The Teranet index displays a 5\% year-on-year house price
increase throughout 2003. Price increases remain strong in subsequent
years, reaching 7-9\% between October 2007 and May 2008. This may
be due to the relatively limited number of SARS cases in Toronto.
This is also consistent with a model in which house prices capitalize
the entire flow of future rents and thus are resilient in the face
of short-run shocks such as pandemics.

Overall, the results described in this chapter imply that metropolitan
areas may be on an equilibrium path, and that shocks are short run
deviation from this single dynamic equilibrium. \citeasnoun{davis2008search}
finds no evidence of multiple equilibria in cities' dynamics, using
data for 114 Japanese cities. This paper finds that industrial composition
and the size of the manufacturing sector are unchanged after large
shocks to city population and employment. This resilience may be due
to the quality of institutions of high-income countries~\cite{kahn2005death}. 

This chapter proceeds as follows. Section~\ref{sec:Two-shocks} describes
the ongoing shocks experienced by U.S. housing markets since March
2020, and their impact on market dynamics. Section~\ref{sec:Looking-Forward}
describes long-run 1970\textendash 2010 evidence of the drivers of
metropolitan population growth, as well as new evidence of the impact
of September 11, and the 1989 Loma Prieta earthquake on neighborhood
dynamics. Section~\ref{sec:Conclusion} provides a cautious forecast
of urban resilience in the face of the 2020 shocks.

\section{The U.S. Housing Market in 2020: Resilience and Adaptation\label{sec:Two-shocks}}

\subsection{Short-Run Aggregate Dynamics: Insights from Prices and Rents}

Time series of house prices, listings, and rents suggests that 2020
is a major shock to real estate dynamics. It also provides evidence
about the market's expectation of resilience. We describe the dynamics
using Zillow's time series data and interpret them using standard
principles of real estate: while rents reveal the current flow value
of housing, prices capitalize current and future flow values. This
section uses metro-level time series. In the next section we focus
on smaller, more granular, local housing market dynamics at the 5-digit
ZIP code level. 

\begin{figure}
\caption{The U.S. Housing Market in 2020: Aggregate Dynamics}

\emph{Panels (a), (b), (c) provide simple statistics on year-on-year
changes in house values, rents, and inventories for the US (bold line)
and for the two largest metropolitan areas (dotted and dashed lines).
Inventories are not available for the same time period as prices.
Panel (d) presents two histograms of price changes in red (resp. rent
changes in blue) for the 100 largest metropolitan areas.}

\begin{center}

\subfloat[YoY Changes in Estimated House Values\label{fig:yoy_changes_values}]{

\includegraphics[scale=0.45,trim={0cm 0cm 0cm 2cm},clip]{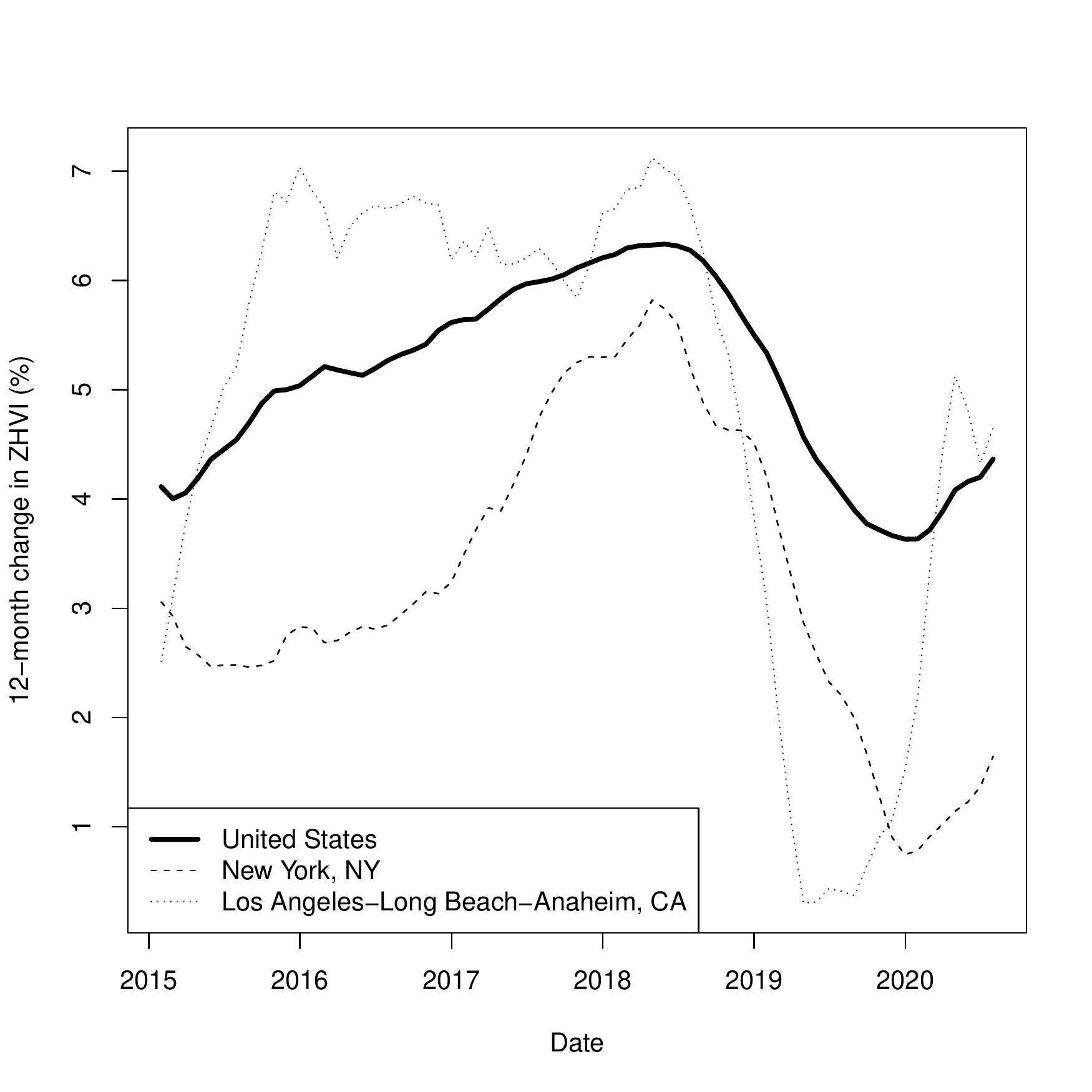}}\subfloat[YoY Changes in Estimated Rents\label{fig:YoY-Changes-in-Rents}]{

\includegraphics[scale=0.45,trim={0cm 0cm 0cm 2cm},clip]{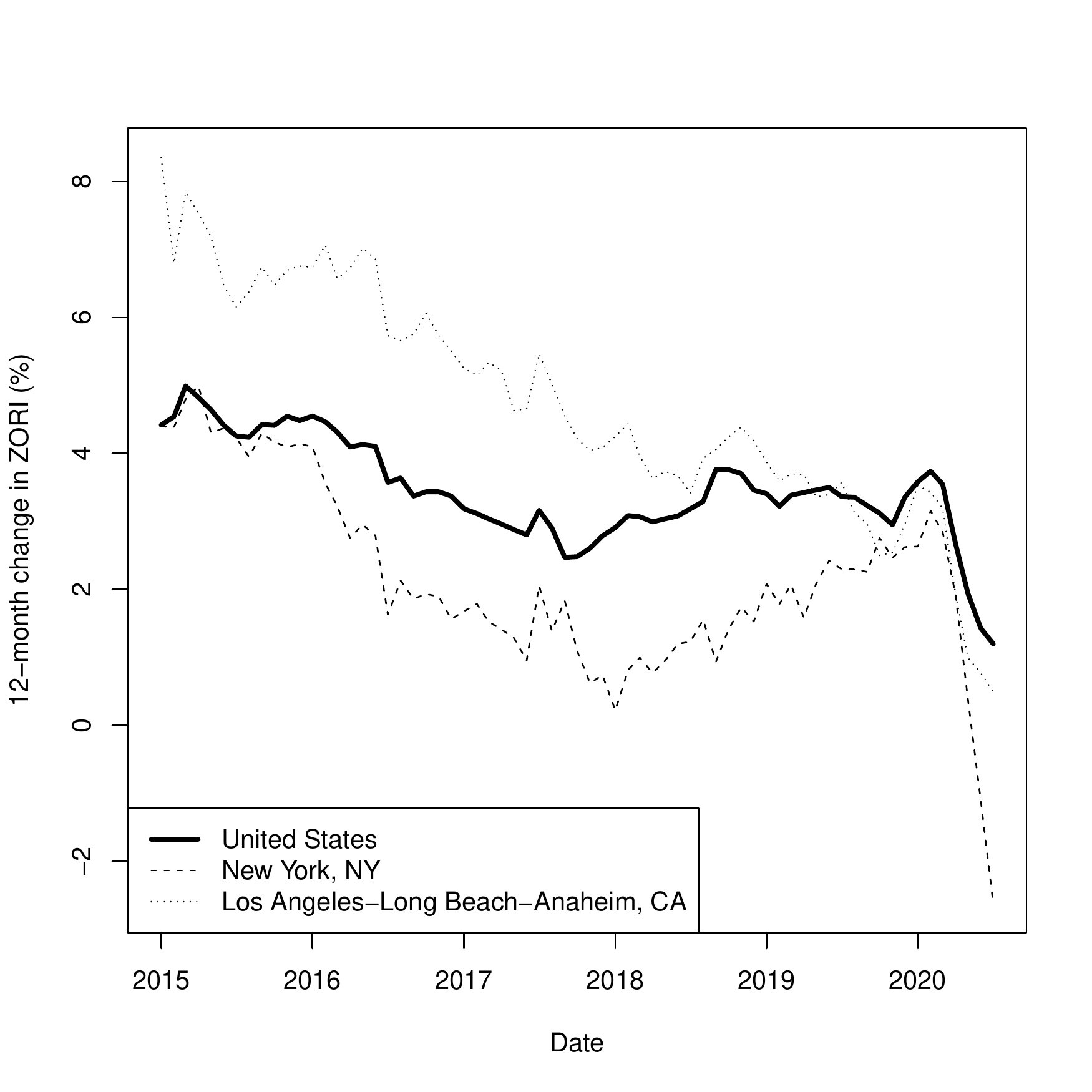}}

\subfloat[YoY Changes in Inventories\label{fig:YoY-Changes-in-Inventories}]{

\includegraphics[scale=0.45,trim={0cm 0cm 0cm 2cm},clip]{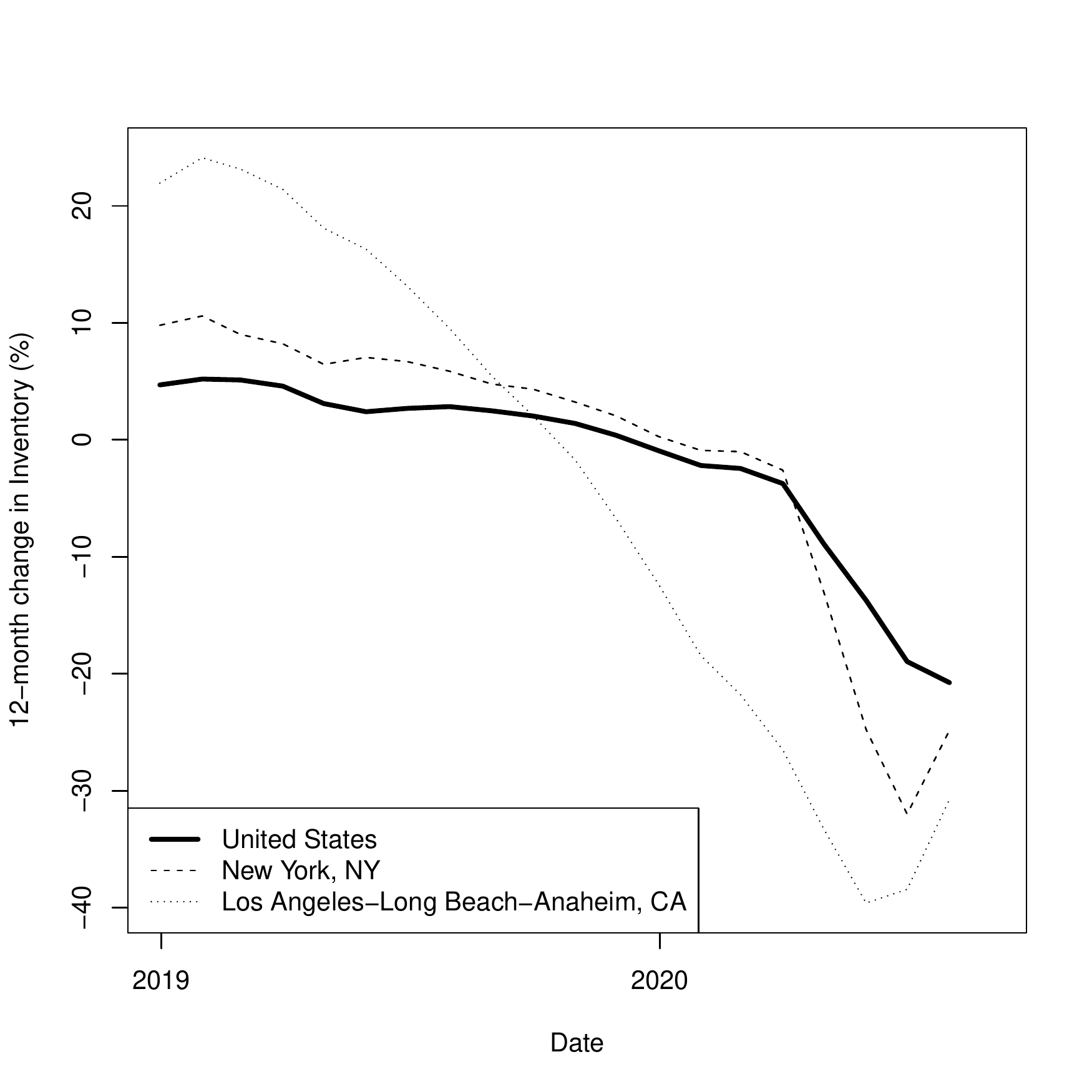}}~~~~~\subfloat[Distribution of Price Changes Across Metro Areas, March to August
2020\label{fig:Distribution-of-Price-and-Rent-Changes}]{

\includegraphics[scale=0.45]{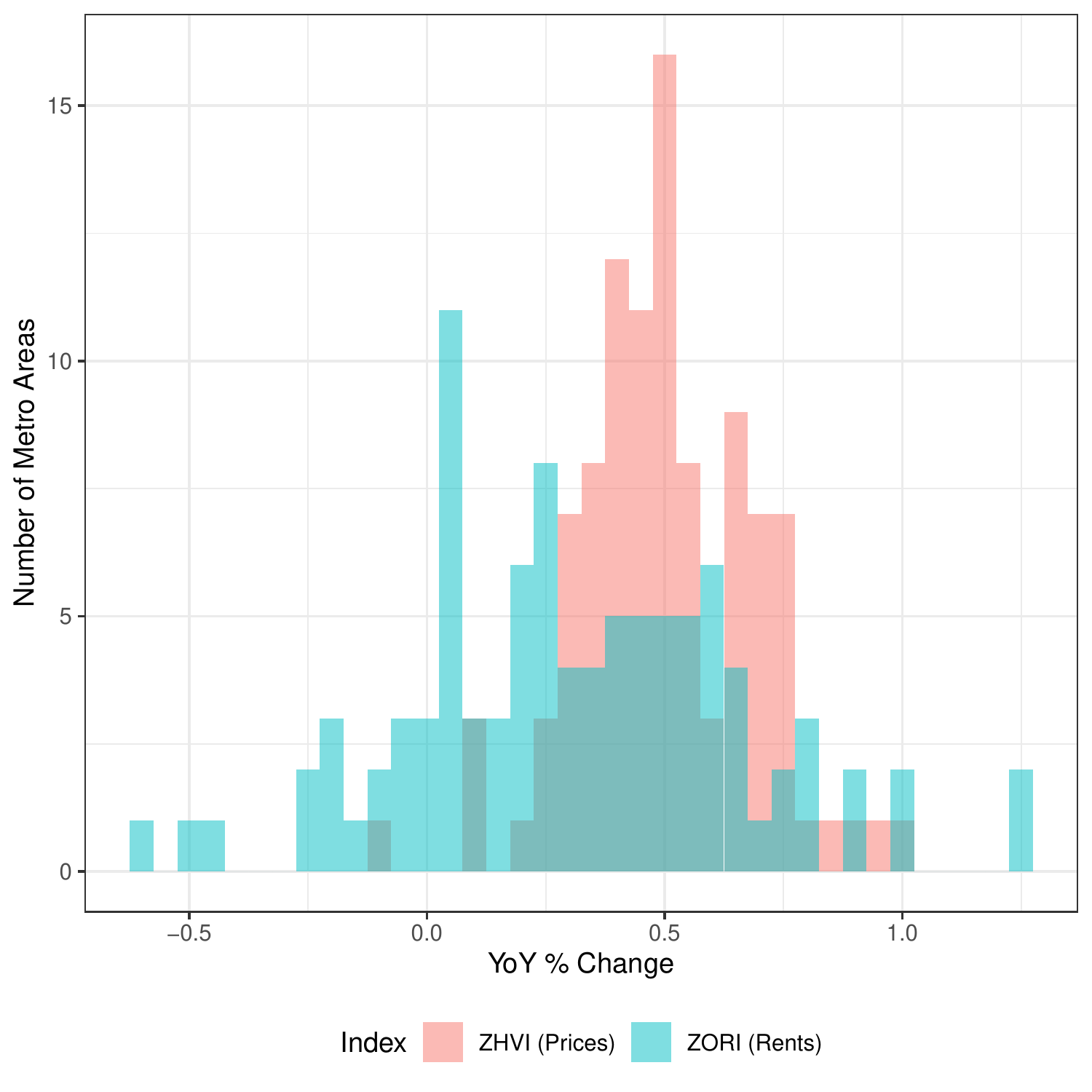}}

\end{center}
\end{figure}

Figure~\ref{fig:yoy_changes_values} presents the year-on-year change
in the Zillow House Value Index (ZHVI) between January 31, 2015 and
July 3, 2020. This price index is built using a repeat-sales methodology
similar to \citeasnoun{case1987prices}. The bold line is for the
United States, the dashed line for the tristate metropolitan area
of New York; and the dotted line for the metropolitan area of Los
Angeles. All three series suggest that after a deceleration of prices
in 2019, transaction prices experienced an accelerating growth from
January till July 2020. Perhaps surprisingly, such deceleration did
not soften during the Covid-19 pandemic, but rather price growth accelerated,
reaching year-on-year levels above 4\% in July.

The dynamic of rents is rather different, and reconciling this apparently
contradictory dynamics provides new insights. Figure~\ref{fig:YoY-Changes-in-Rents}
displays the year-on-year change in the Zillow Observed Rent Index
(ZORI), which measures changes in asking rents over a sample of properties.
By measuring rents for the same units, this index is akin to a ``repeat-rent''
index. Hence this index is built with a similar method as the house
value index. The figure suggests that rent growth not only decelerated,
but rents decreased in the metro area of New York, dropping by more
than 2\% year-on-year in July 2020. In the United States overall and
in Los Angeles, rents are close to declining. Figure~\ref{fig:Distribution-of-Price-and-Rent-Changes}
presents the metro-level distribution of average year-on-year changes
during the March to August 2020 period, for the 100 largest metropolitan
areas. It suggests that overall prices have increased faster than
rents, with a significant number of metropolitan areas experiencing
rent declines or stagnation; while there is only one metro area with
price declines. 

Listings experience the largest drop. Figure~\ref{fig:YoY-Changes-in-Inventories}
suggests that listings started decreasing significantly at the beginning
of the pandemic, dropping year-on-year by 20\% in the U.S. and by
up to 40\% in Los Angeles, with a rebound in June-July 2020. This
presents a first hypothesis for the seemingly paradoxical increase
in prices. A first hypothesis is selection bias. Houses that do not
transact during a given time period do not contribute to a repeat
sales index by construction, and houses in the lower part of the price
distribution are more likely to experience no transaction during downturns~\cite{ouazad2019market}.
While houses that do transact experience price increases, houses whose
value is declining might not contribute to the set of observations
of the price index. Hence part of the index's fluctuations may simply
be due to dynamic selection~\cite{gatzlaff1997sample}. This possibility
nevertheless is unlikely to explain the observed price and rent trends.
First, both the price index and the rent index are vulnerable to this
selection bias. Second, one econometric approach to correcting for
such selection bias, the inverse time weighting approach~\cite{ambrose2015repeat},
does not typically yield significant differences in the price index.

Three alternative mechanisms rationalize the evolution of the housing
market's trends. The simplest way to express them is using the \citeasnoun{gordon1956capital}
approach. Such approach capitalizes expected rents using a constant
discount factor and a constant expected growth rate of rents. Rents
are net of maintenance costs, property taxes, and potential credit
costs. Formally,
\begin{equation}
\frac{R}{P}=r-g,\label{eq:gordon}
\end{equation}
where $R$ is the current net rent, $P$ the current value of the
asset, $r$ the required capital yield, and $g$ the growth rate of
net rents. This can be written as $P=\frac{R}{r-g}$, suggesting that
prices may increase even as \emph{current }rents fall whenever (i)~the
expected growth rate of rents increases, (ii)~the rate of return
$r$ declines, (iii)~net rents increase relative to gross rents due,
for instance, to a decline in credit costs. 

\begin{figure}
\caption{The U.S. Housing Market in 2020: Explaining the Resilience of Prices\label{fig:Expectations-of-Resilience}}

\emph{These graphs describe the decline in the rent-to-price ratio,
net of maintenance costs and property taxes (figure (a)), and three
key components of the Shapiro-Gordon valuation formula: (b) the 30-year
fixed rate mortgage average, which measures credit costs and affects
net rental yields; (c) the AAA corporate bond yield, a proxy for the
yield on capital; and (d) expectations of rent growth.}

\begin{center}

\subfloat[Estimated Net Rent-to-Price Ratio\label{fig:Estimated-Net-Rent-to-Price}]{

\includegraphics[scale=0.45]{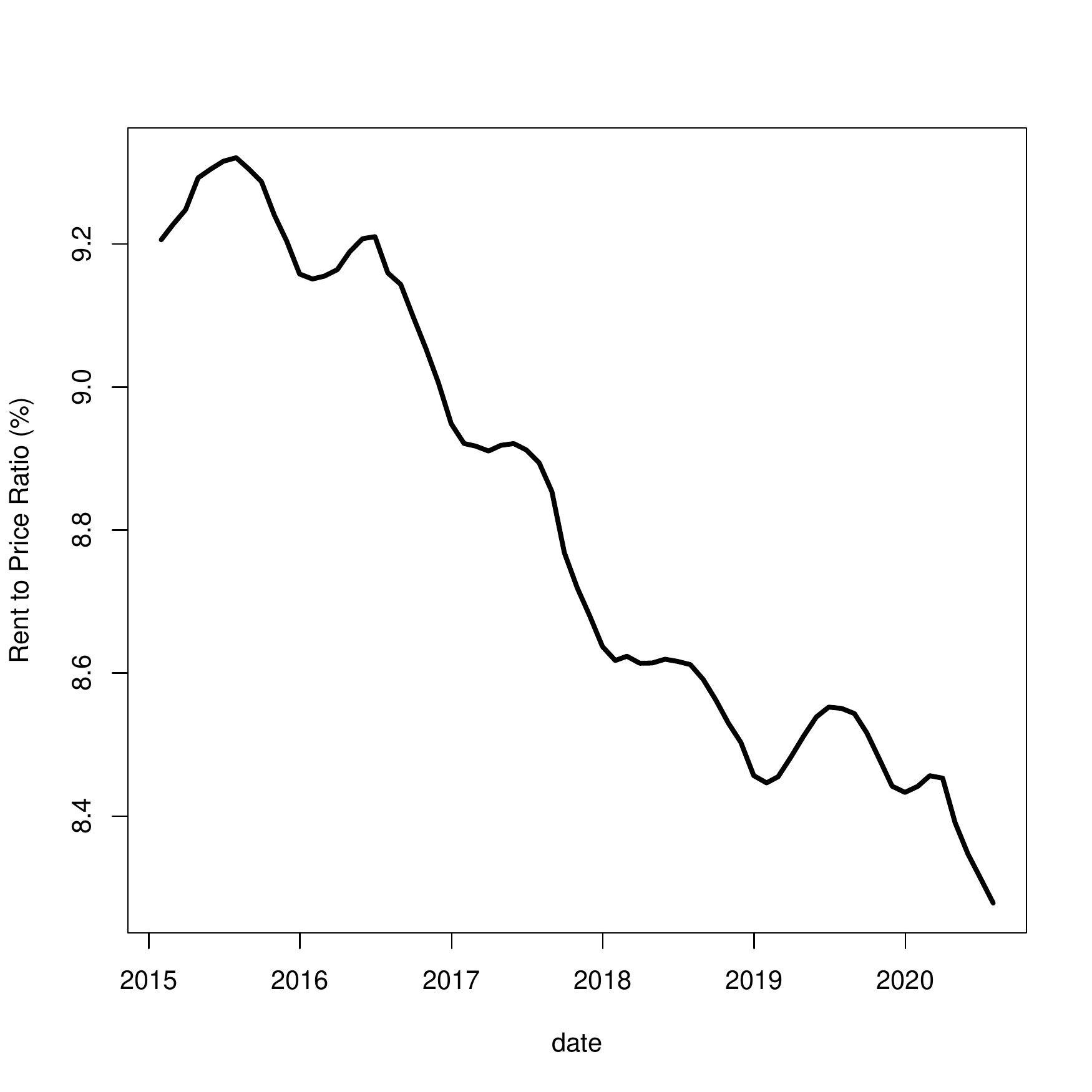}}\subfloat[30-Year Fixed Rate Mortgage Average\label{fig:30-Year-Fixed-Rate}]{

\includegraphics[scale=0.45]{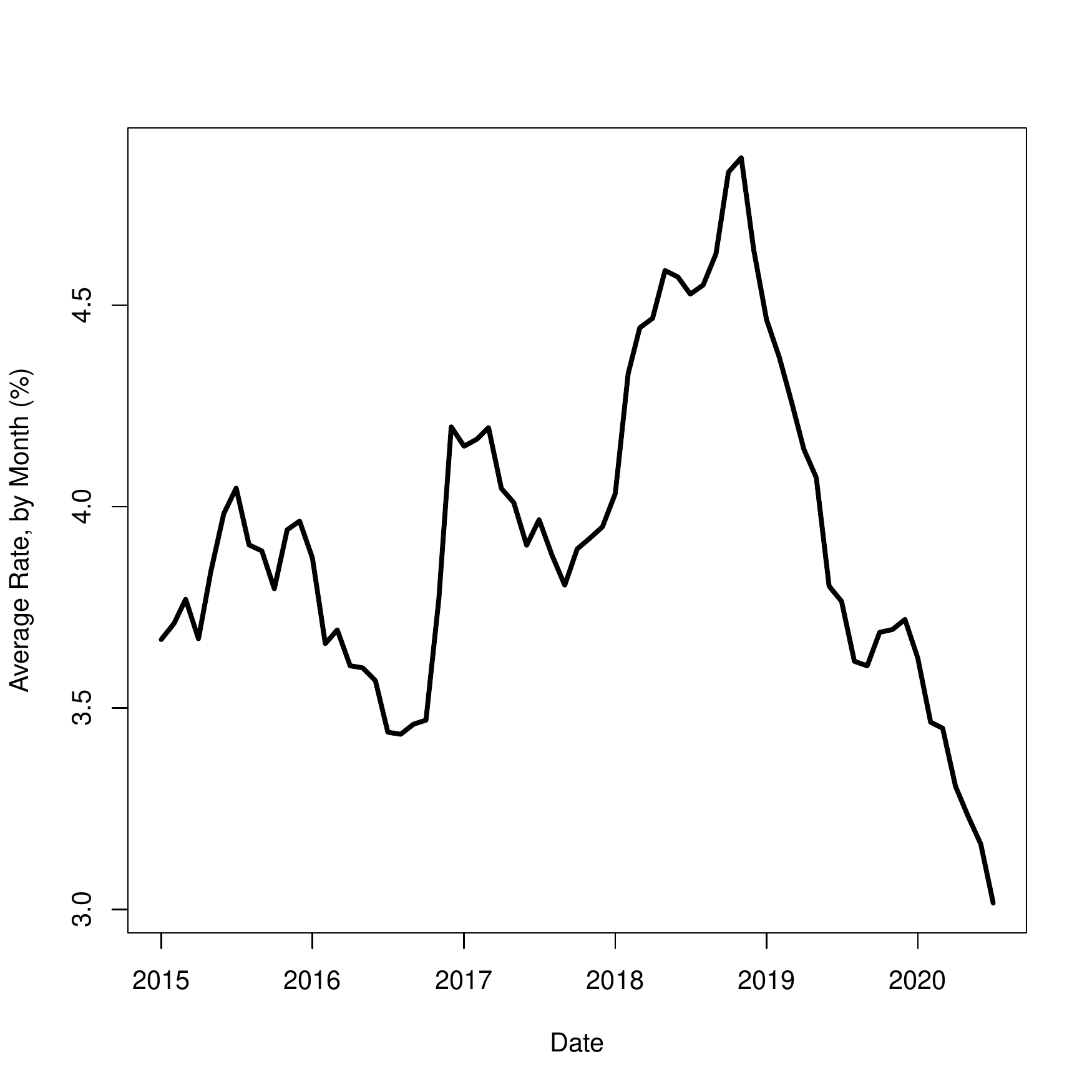}}

\subfloat[Moody's Seasoned AAA Corporate Bond Yield\label{fig:Moody's-Seasoned-AAA}]{

\includegraphics[scale=0.45]{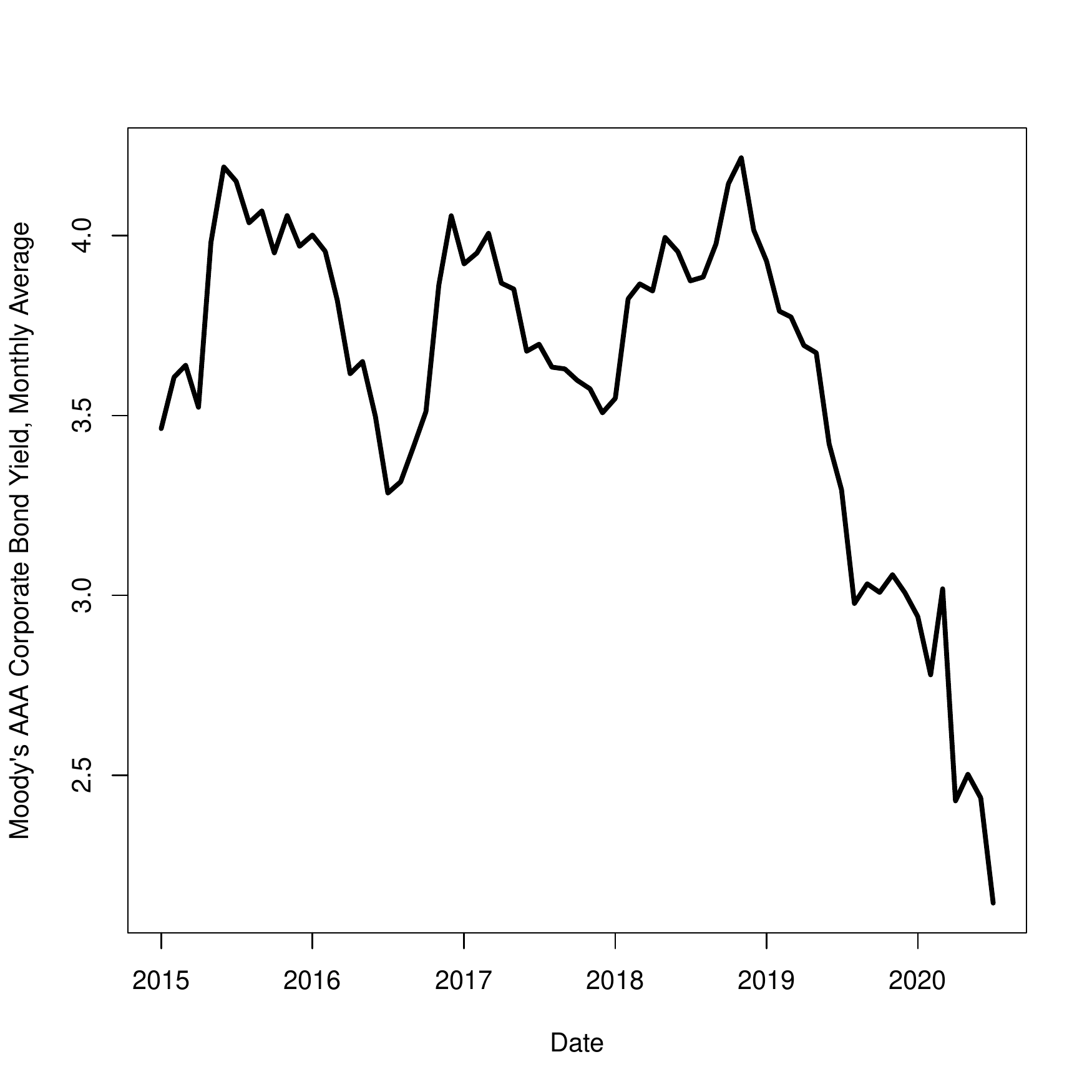}}\subfloat[Expectations of Rent and Price Growth\label{fig:Expectations-of-Rent}]{

\includegraphics[scale=0.45]{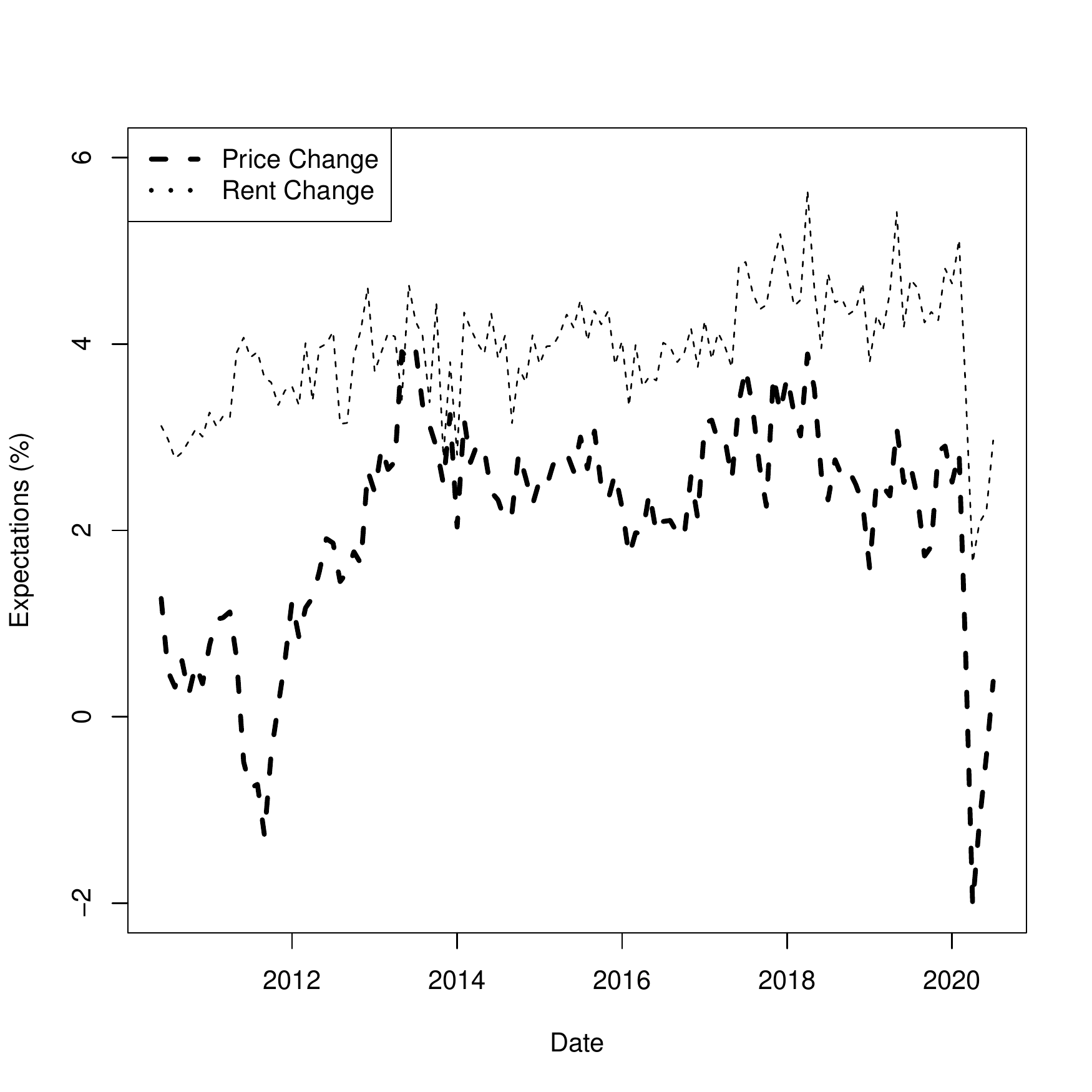}}

\end{center}

\emph{Sources: Zillow ZHVI and ZORI for the rent-to-price ratio. Maintenance
cost from \citeasnoun{harding2007depreciation}. Average property
tax rate from \citeasnoun{malm2015comments}. Federal Reserve of St
Louis series DAAA and MORTGAGE30US. Fannie Mae's July 2020 National
Housing Survey.}
\end{figure}

Figure~\ref{fig:Estimated-Net-Rent-to-Price} shows that as expected,
the increase in prices and the decline of rents implies a declining
rent-to-price ratio, which is the outcome of at least these three
potential mechanisms. First, the decline in the 30-year fixed rate
mortgage average (Figure~\ref{fig:30-Year-Fixed-Rate}) lowers interest
costs and pushes prices up at given rents. The impact of cheap mortgage
credit on house prices has been documented~\cite{adelino2012credit,favara2015credit,justiniano2019credit}.
Second, the decline in the AAA corporate bond yield (Figure~\ref{fig:Moody's-Seasoned-AAA})
suggests that prices are increasing in a search for yields. The required
rate of return on capital can be approximated by such a safe bond
yield plus a risk premium. Third, the increase in prices and the decline
of the price-to-rent ratio is consistent with expectations of rent
growth; while current rents may be low, buyers are arguably expecting
substantial \emph{future }rent growth. Figure~\ref{fig:Expectations-of-Rent}
plots expectations of price and rent growth using the time series
of Fannie Mae's National Housing Survey. While expectations of rent
growth ($g$) fall sharply in June 2020, they rebound and become positive
again in July 2020, suggesting that housing market participants expect
a short-lived trough in rents rather than a prolonged slowdown.

\subsection{Covid-19 Cases:\protect \\
Greater Frictions, Declining Rents, Resilient Prices}

The global Covid-19 pandemic affected housing markets throughout the
world. Yet, the spatial distribution of confirmed cases and deaths
is uneven. The pandemic emerged in the United States as a significant
measurable phenomenon in the first half of March 2020. While daily
confirmed cases were below 70 a day on March 5th, they grew to a peak
of more than 77,000 cases a day on July 16th 2020 for a total of 5.9
million cases as of August 29, 2020.\footnote{This chapter was written in September 2020.}
On the same day, Canada had reached a total confirmed number of cases
of more than 129,000 cases. 

\begin{figure}
\caption{The U.S. Housing Market in 2020: Covid-19 Infections and the Housing
Market}

\begin{center}

\subfloat[Confirmed Covid-19 Cases Per Capita across Metro Areas\label{fig:Covid-19-Infections-and-urban-density}]{

\includegraphics[scale=0.12,trim={5cm 0cm 5cm 0cm},clip]{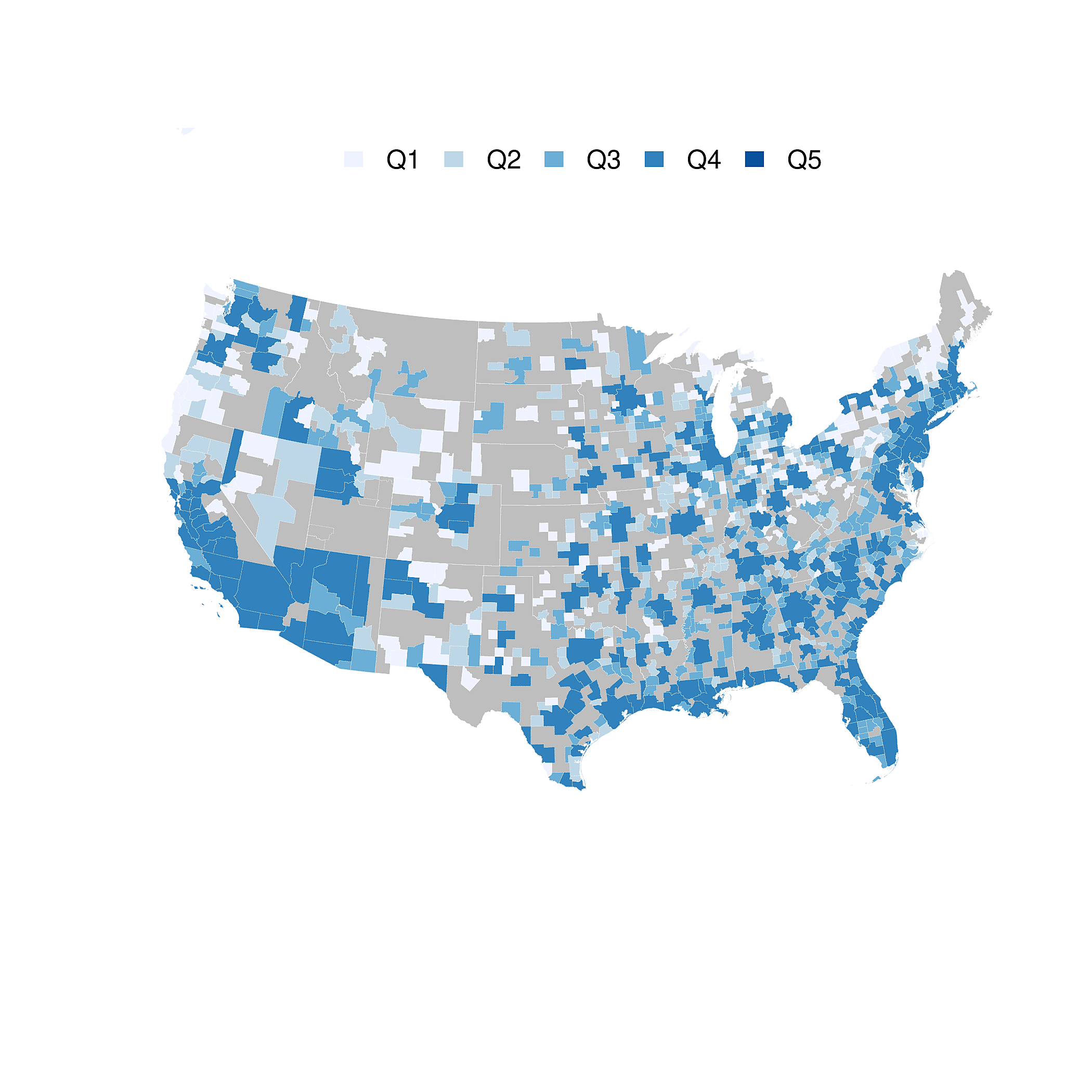}}~~~~\subfloat[Covid-19 Cases and Inventories\label{fig:Covid-19-Infections-and-urban-density-1}]{

\includegraphics[scale=0.45]{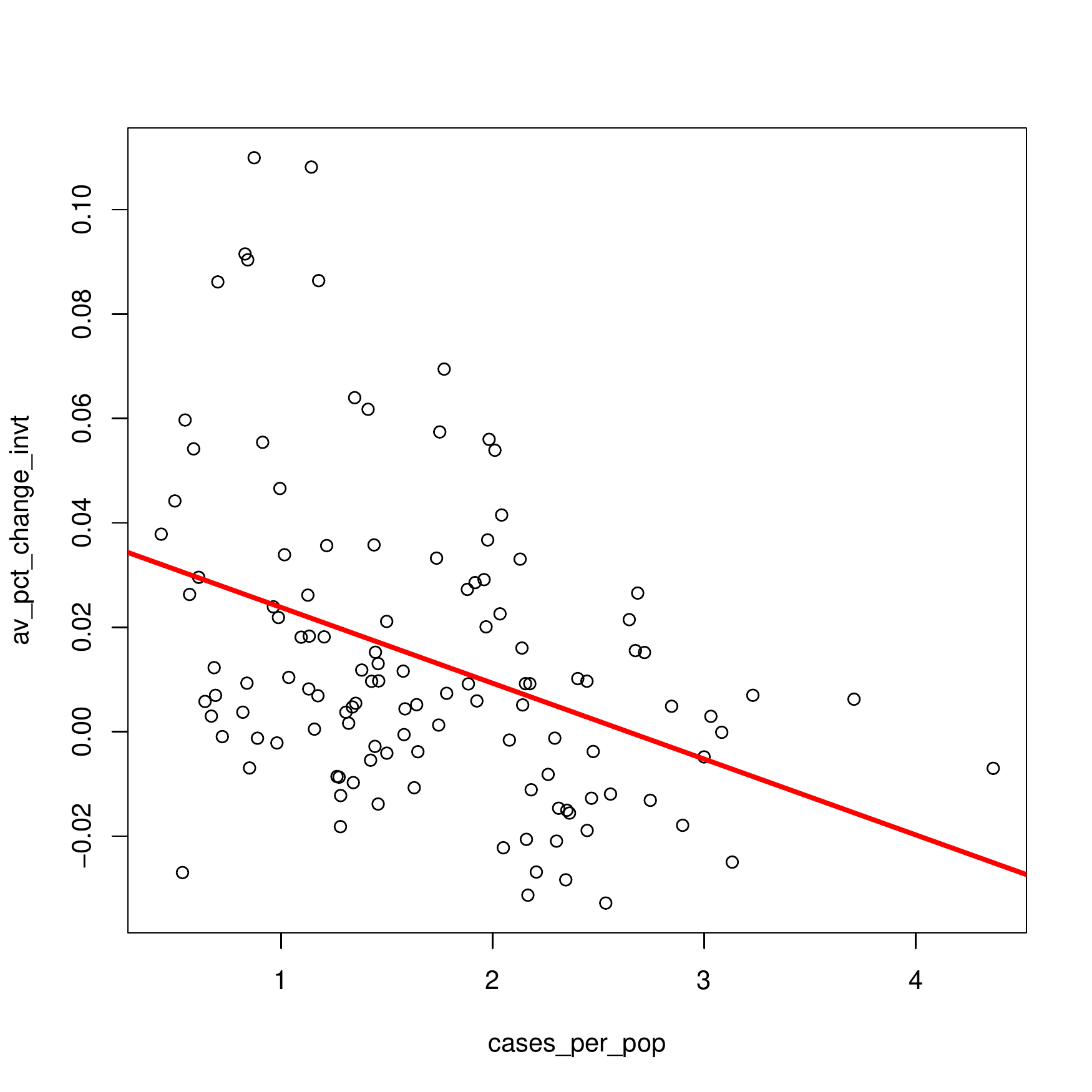}}

\subfloat[Covid-19 Cases and House Prices\label{fig:Covid-19-Infections-and-prices}]{

\includegraphics[scale=0.45]{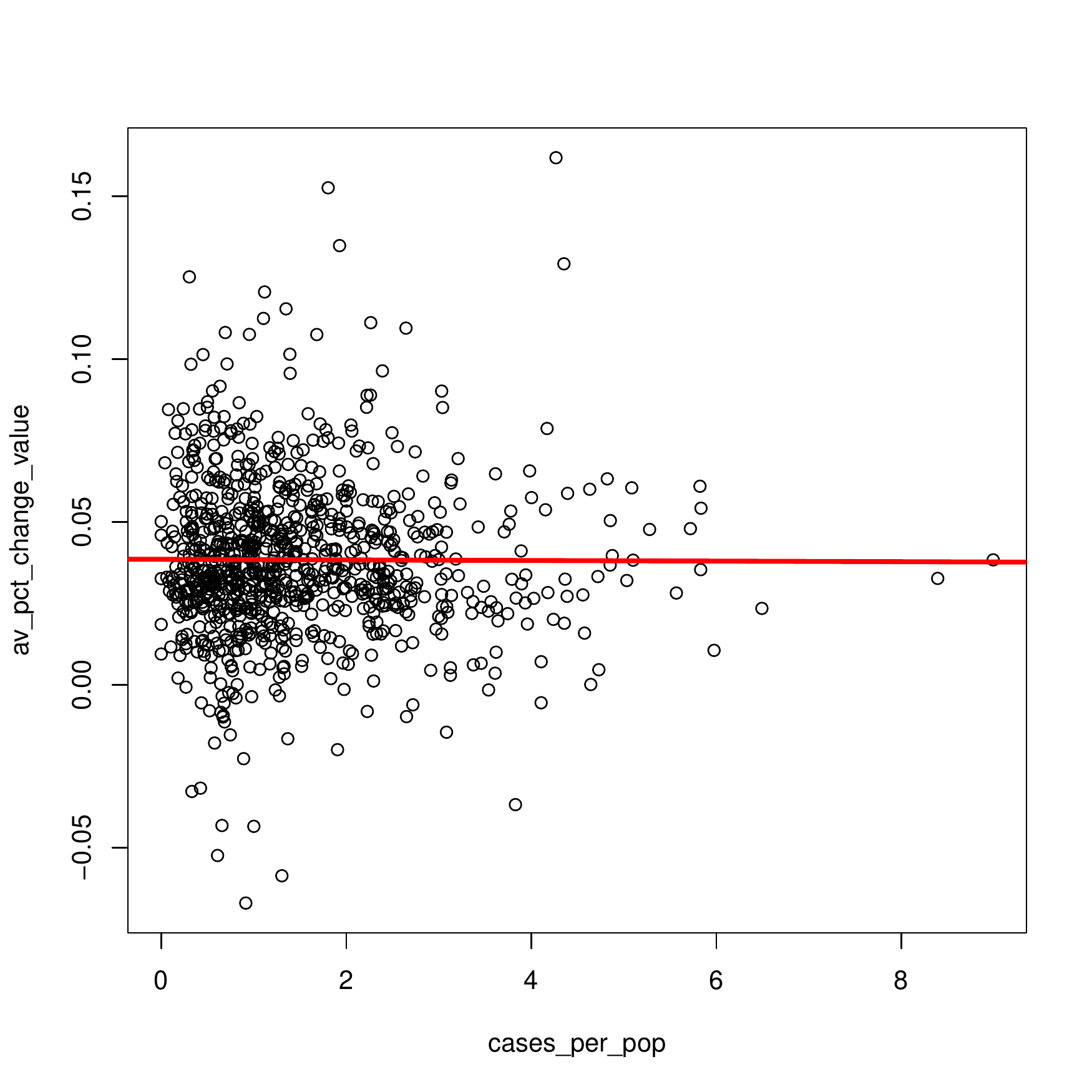}}~~~~\subfloat[Covid-19 Cases and Rents\label{fig:Covid-19-Infections-and-rents}]{

\includegraphics[scale=0.45]{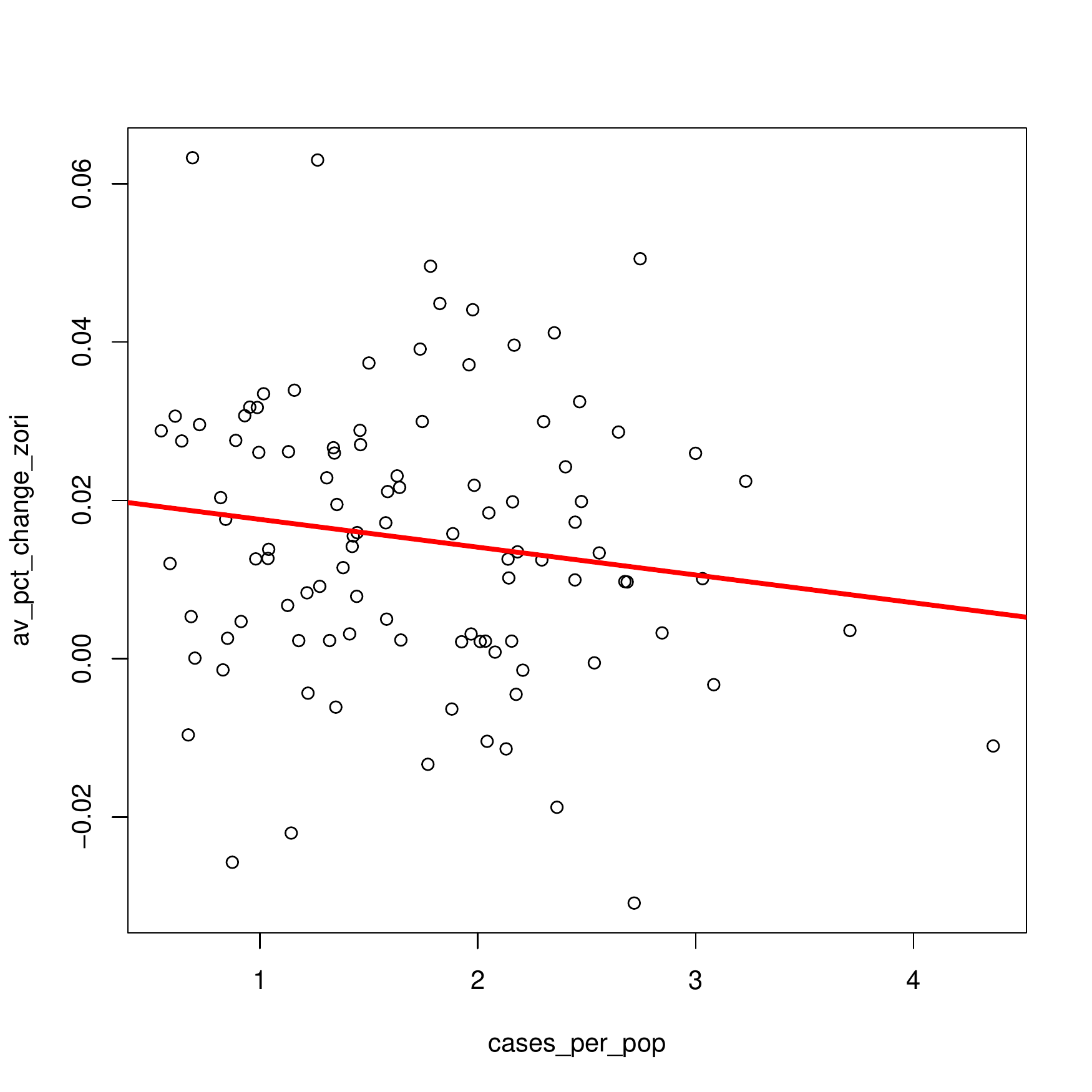}}

\end{center}
\end{figure}

Figure~\ref{fig:Covid-19-Infections-and-urban-density} presents
the spatial distribution of cases per capita across Zillow's metropolitan
housing markets. The colors corresponds to quantiles of cases per
capita. This map suggests that Covid-19 infections reached most housing
markets, with an average number of confirmed cases per 100 residents
of between no confirmed case (three metros of Utah: Cedar City, Price,
and Saint George) and a maximum of 9 cases per 100 residents (Alta,
Indiana). As expected, the largest metropolitan areas host the largest
number of total confirmed cases, with 543,000 cases in New York, 282,476
in Los Angeles. With the exception of Riverside, California, the largest
numbers of cases are all in the 10 largest metropolitan areas by population. 

Table~\ref{tab:regression_drivers_covid19_infections} performs a
county-level regression of confirmed cases per capita on a range of
variables from the American Community Survey. Density is measured
by the ratio of county population on the county's area in squared
kilometers. The log density is a more relevant measure than density
itself as the regression is less driven by extreme observations. The
regression includes state fixed effects \textendash{} results are
unaffected by the inclusion of state fixed effects. The table displays
an economically and statistically significant correlation between
county log density and confirmed cases per capita regardless of the
inclusion of additional controls.

We match such cases by population to shifts in inventories to document
a substantial and significant correlation between the decline in real
estate inventories and the number of cases per population. This is
depicted in Figure~\ref{fig:Covid-19-Infections-and-urban-density-1}.
The vertical axis is the average year-on-year percentage change in
inventories over the March to August 2020 period. The horizontal axis
is the number of cases per population, where the total number of confirmed
cases is from the Johns Hopkins Coronavirus Research Center; and county-level
population aggregated to Zillow's metro areas is from the 2018 American
Community Survey.\footnote{While 2020 county-level population numbers have not yet been released,
a similar correlation would arguably hold with updated data.} The pandemic affected the ability of homeowners to sell and of buyers
to acquire a property, likely increasing search frictions and leading
to inefficiencies. There is no metropolitan area with cases per population
above the median \emph{and} inventory growth above the median. Charleston,
South Carolina, with more than 3 cases per 100,\footnote{Confirmed cases are also reported as cases per million. Using this
alternative scaling does not affect this chapter's analysis.} experienced a 4\% decline in inventories. New Orleans, with 3.1 cases
per 100, experienced a 3.7\% decline in inventories. In contrast,
some of the largest increases in inventories happened in metropolitan
areas with low case numbers: San Francisco, with only 1.1 cases per
100, experienced a +3.2\% increase in listings.

There is no detectable metro-level correlation between house price
dynamics and the number of confirmed cases, suggesting that the impact
of the pandemic may be more likely to stem from the economic consequences
of the pandemic rather than through the avoidance of infection probabilities.
Figure~\ref{fig:Covid-19-Infections-and-prices} plots the average
monthly change in prices for each of the largest 100 metros against
the number of cases per population. It suggests that prices are largely
unrelated to confirmed cases, with a large variance of up to 20 percentage
points, in house price changes for metros with low infection numbers.
And no significant difference between metros with low infection numbers
and metros with high infection numbers.

Evidence may come from the correlation between rents and infection
numbers. Figure~\ref{fig:Covid-19-Infections-and-rents} plots the
average change in rents against the number of cases per population.
Metro areas with large numbers of cases per population experienced
lower than average rent growth. In contrast, metropolitan areas with
low case counts per population experienced some of the largest rent
growth levels.

These three pieces of evidence (on inventories, prices, rents) suggest
a substantial short-run impact of the pandemic on the flow utility
of housing in metro areas affected by the pandemic, but a long-run
expectation of resilience whereby the pandemic does not significantly
affect buyers' expectations of the value of living in metro areas
with large cases per population. 

\subsection{Evidence of Short-Run Suburbanization}

While house prices are overall on the rise, there may be \emph{within}-\emph{city}
shifts in demand towards neighborhoods that are less dense and farther
away from the central business district, which are arguably less exposed
to the pandemic. Anecdotal evidence\footnote{``New Yorkers Look To Suburbs And Beyond. Other City Dwellers May
Be Next'', National Public Radio, July 8, 2020. ``New Yorkers Are
Fleeing to the Suburbs: \textquoteleft The Demand Is Insane\textquoteright ``,
New York Times, August 30, 2020.} suggests that cities may become more resilient when households increase
their demand for less dense areas where the propensity for infections
is perceived to be lower.\footnote{While many other factors than density explains the variance of cases
across locations, there is a significant and positive correlation
between population density and cases per capita, as displayed in Table~\ref{tab:regression_drivers_covid19_infections}.} To perform this analysis, we turn to neighborhood-level evidence
from the New York City metro area.

\begin{figure}
\caption{Low- and High-Population Density ZIP Codes: Two Typical Examples\label{fig:Low--and-High-Population-Density}}

\emph{These two maps present the layout of buildings and roads in
two sample ZIP codes. The ZIP code of the upper panel is part of New
York's Upper East Side, with a high population density of 53,029 residents
per squared kilometers, 18 times that of the ZIP of the lower panel.
Such ZIP code includes the Great Neck Estates on the northern part
of Long Island. It has a population density of 2,968 residents per
squared kilometers. Maps have different scales. }

\begin{center}

\subfloat[Higher Density: The Upper East Side, ZIP 10075]{

\includegraphics[scale=0.65,trim={1cm 3cm 1cm 3cm}, clip]{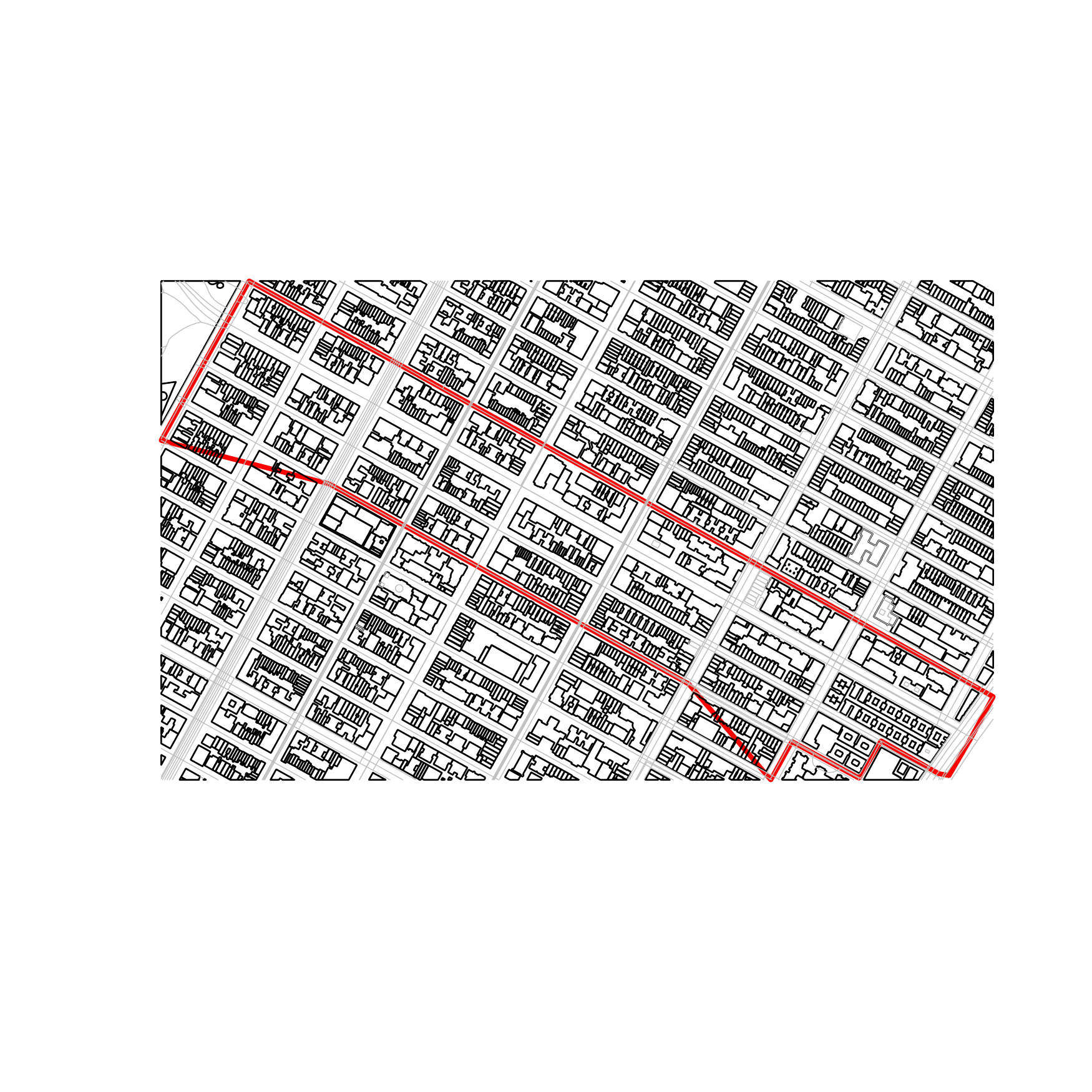}}

\bigskip{}
\subfloat[Lower Density: Russell Gardens, Great Neck Plaza, Great Neck Estates
ZIP 11021]{

\includegraphics[scale=0.65,trim={1cm 3cm 1cm 3cm}, clip]{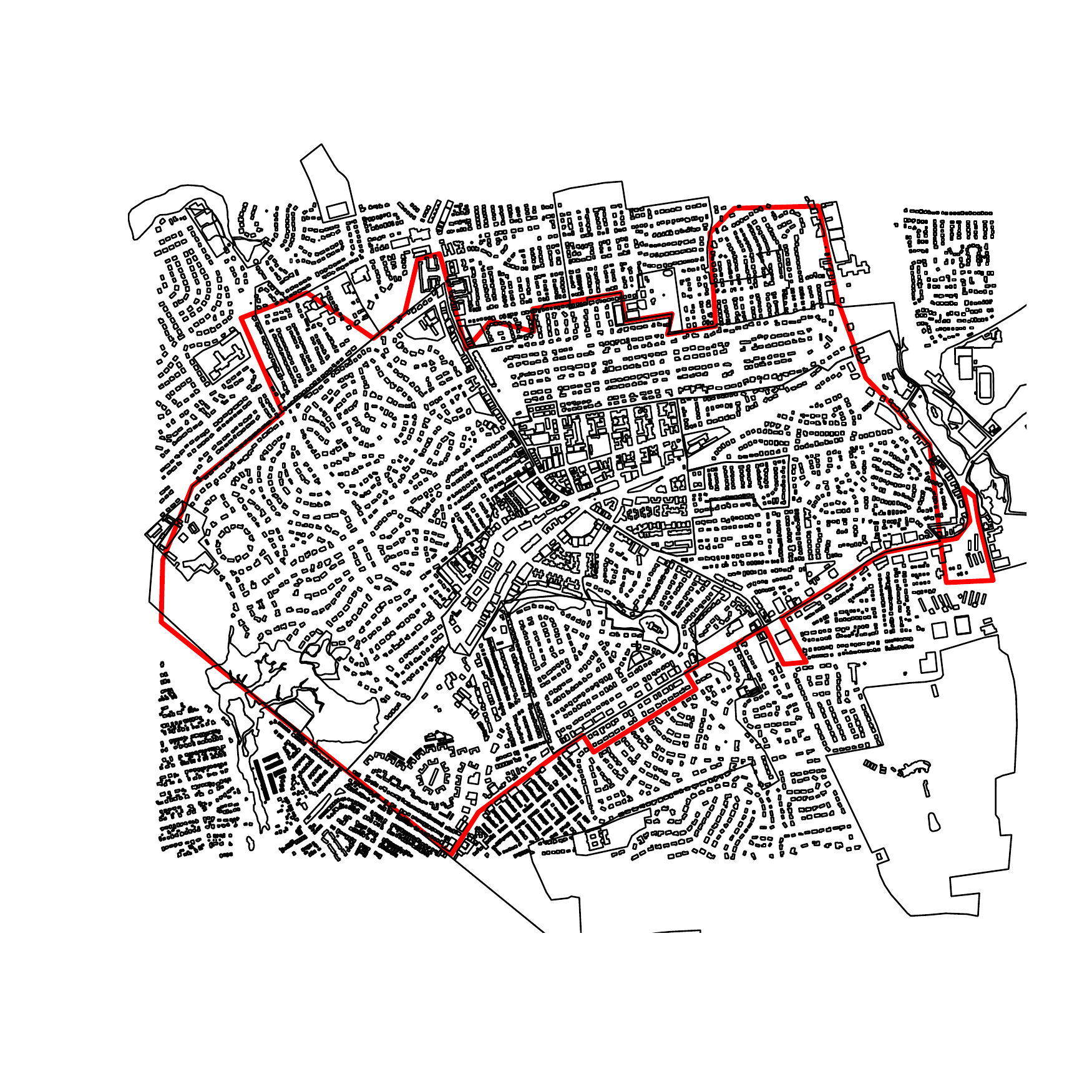}}

\end{center}

\emph{ZIP boundaries projected according to the Census 2010 boundaries.
Building footprint and roads current as of 2020 from Open Street Map.
Population counts from the 5-year averages of the 2018 American Community
Survey.}
\end{figure}

Figure~\ref{fig:Low--and-High-Population-Density} presents the example
of two neighborhoods with two extreme density levels. The upper panel
(a) presents the Upper East Side, with a population density of 53,029
residents per squared kilometer as of 2018. It features condominium
towers and other high density urban developments. Such density is
higher than the average density of the densest cities in the world.
This stands in contrast with New York's Great Neck Peninsula (lower
panel (b)), on the northern side of Long Island, with a population
density 18 times lower, of 2,968 residents per squared kilometer.
While commuting time from Great Neck to downtown Manhattan is less
than half an hour, this neighborhood has more than 20 parks across
9 villages, and features ``verdant residential areas.''\footnote{Marcelle Sussman Fischer, \href{https://www.nytimes.com/2016/07/31/realestate/on-the-great-neck-peninsula-a-rich-blend-of-many-cultures.html}{the New York Times},
July 2016.} 

We test whether neighborhoods such as the Upper East Side have seen
a decline in demand relative to neighborhoods such as Great Neck during
the period of March to August 2020. We do so by regressing shifts
in prices on 1) the distance to the population-weighted center of
the metropolitan area, 2) population density, as the ratio of the
2018 ACS population over the area of the ZIP Code Tabulation Area
in squared kilometers. 

\begin{figure}

\caption{The U.S. Housing Market in 2020: Evidence of Suburbanization\label{fig:Is-Housing-Demand}}

\begin{center}

\subfloat[YoY \% Price Changes and Distance to the Center, 2019]{

\includegraphics[scale=0.45]{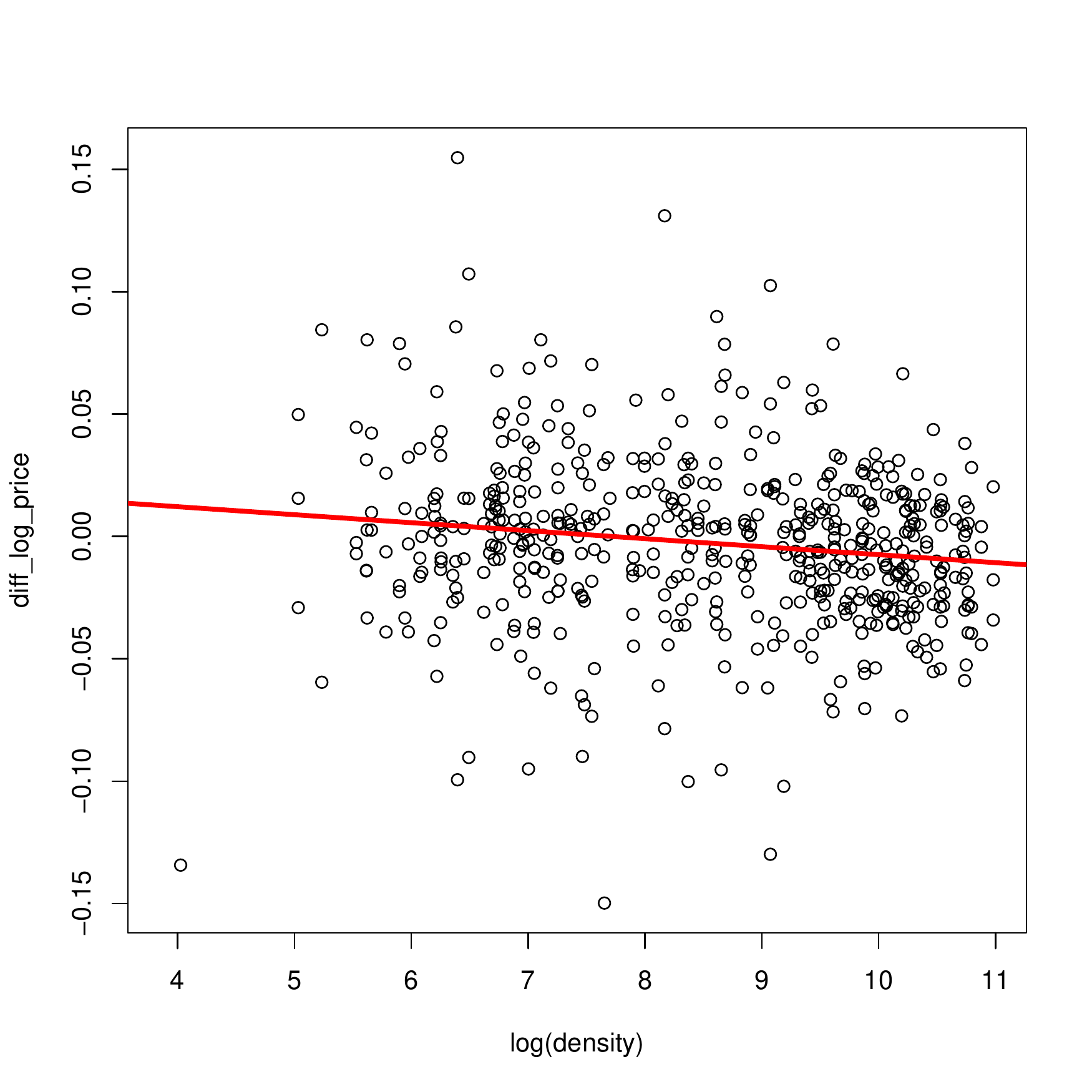}}\subfloat[YoY \% Price Changes and Distance to the Center, 2020]{

\includegraphics[scale=0.45]{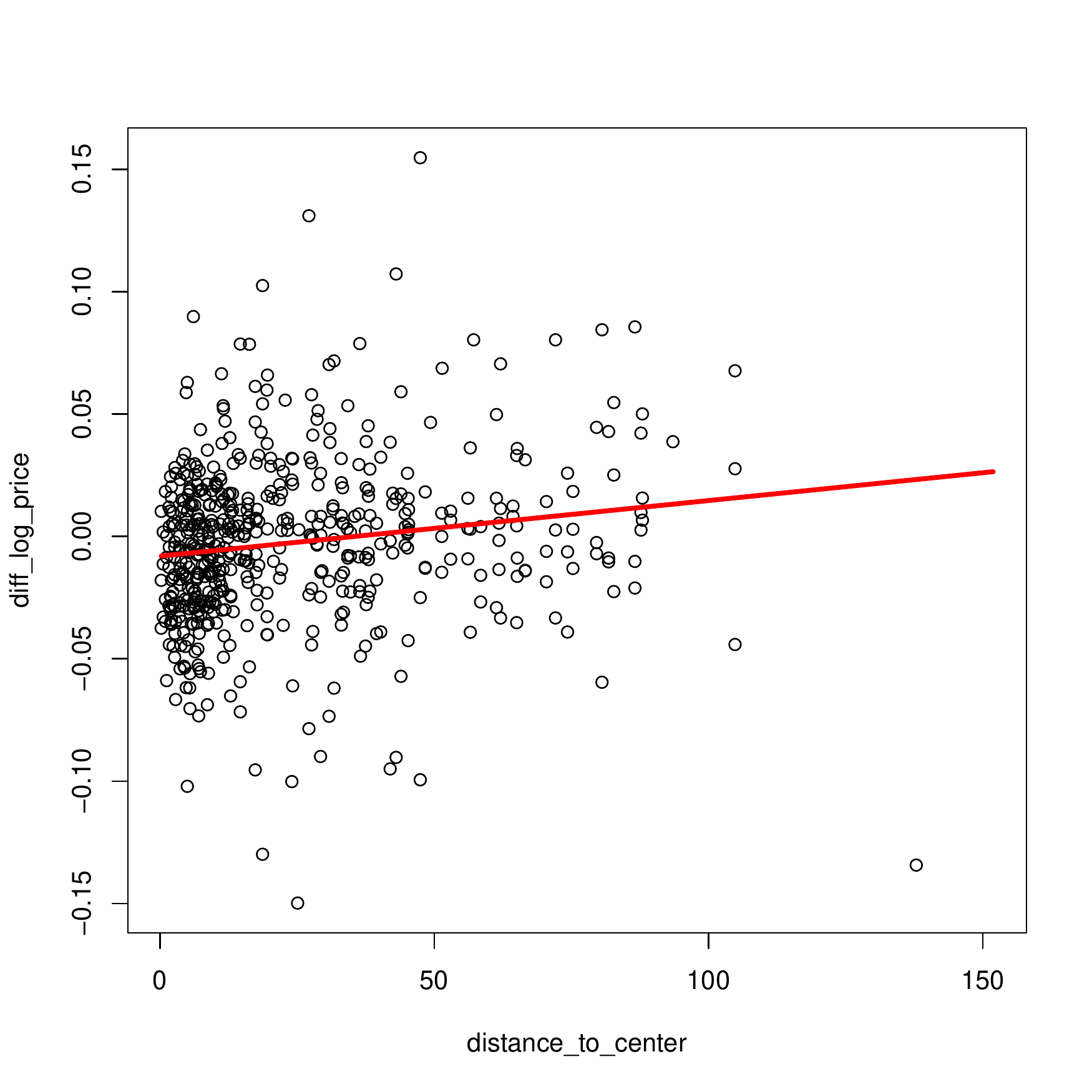}}

\subfloat[YoY \% Price Changes and Density, 2019]{

\includegraphics[scale=0.45]{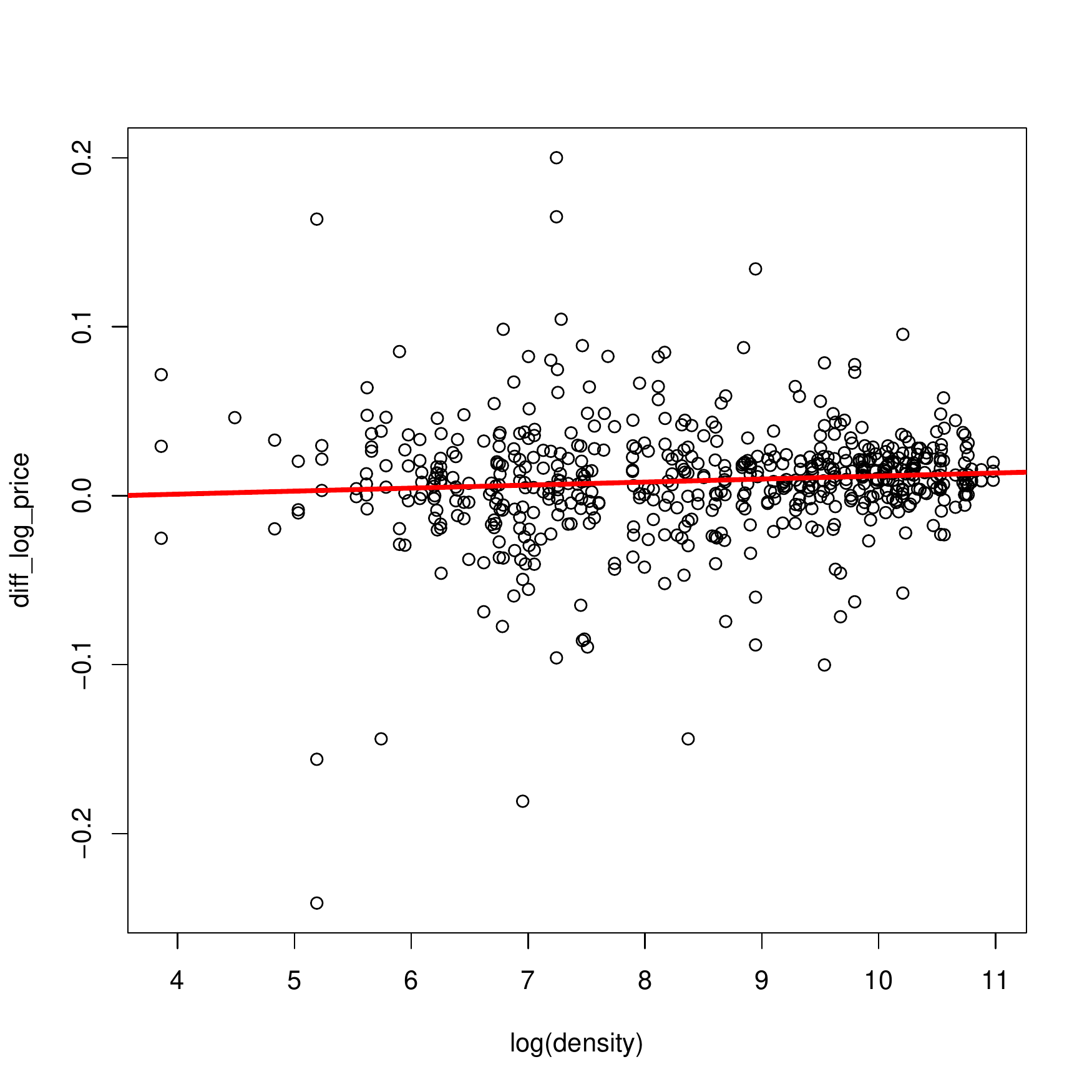}}\subfloat[YoY \% Price Changes and Density, 2020]{

\includegraphics[scale=0.45]{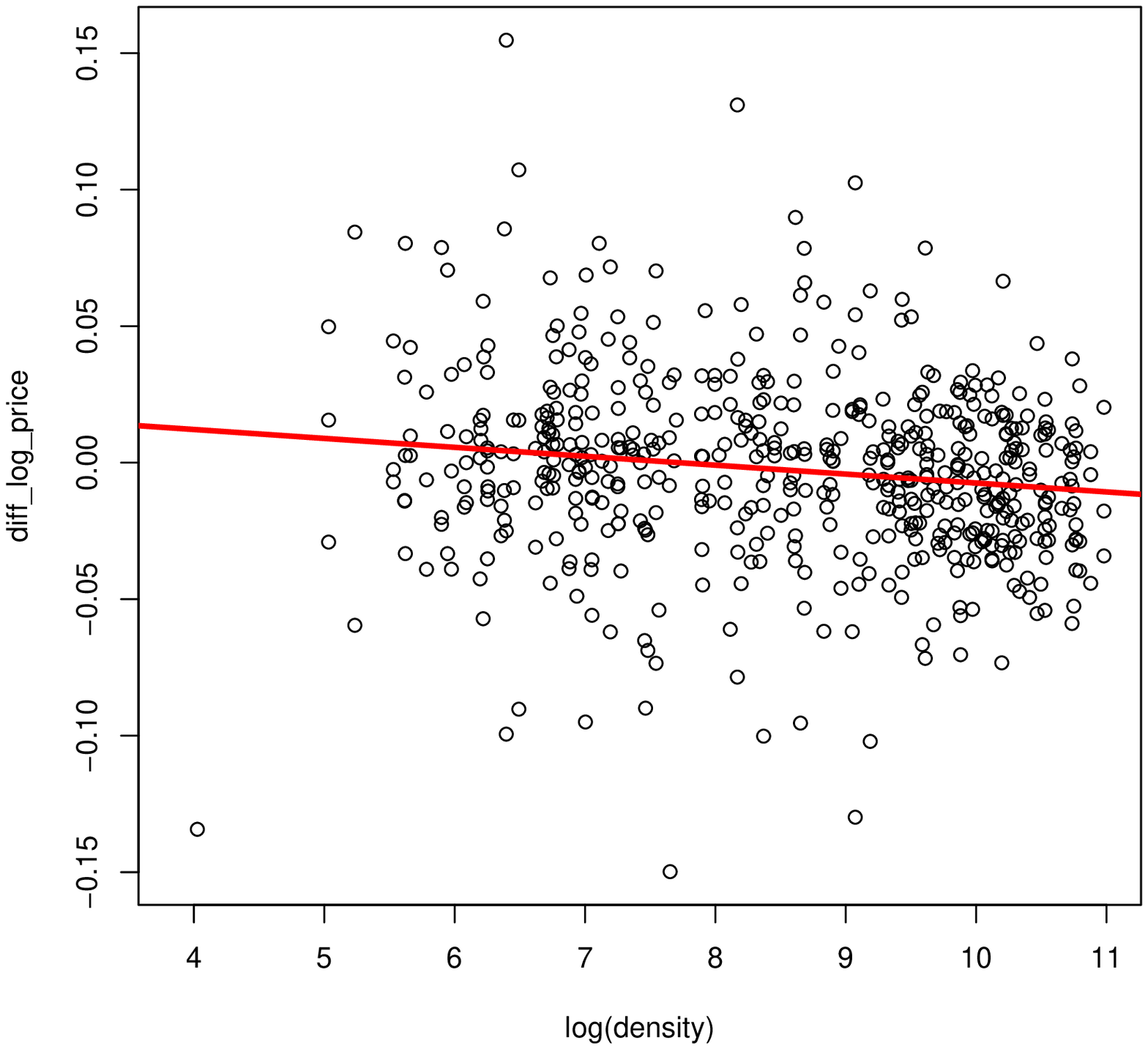}}

\end{center}
\end{figure}

The results are presented in the scatter plot of Figure \ref{fig:Is-Housing-Demand}
and in Table~\ref{tab:Within-City-Adaptation:-Short-Ru}. These scatter
plots and the regression table suggest a \emph{reversal }in patterns
of housing demand during the pandemic. Indeed, the correlation between
house price appreciation and density is positive in the three months
of March-May 2019. That is also true for other periods outside the
pandemic. The correlation between house price appreciation and urban
density is also positive in the same time period of 2019, one year
before the pandemic. Yet these two correlations turn negative and
significant at 1\% in the three months of March to May 2020. As the
supply of housing moves slowly in the short-run, fluctuations in house
prices between March and August are likely a good measure of the shift
in the demand for housing units, vacating less desirable locations,
and searching for housing in more desirable locations. Hence correlations
between the characteristics of neighborhoods and shifts in transaction
prices are likely a relevant proxy for shifts in tastes. These results
suggest that, in the short-run, household demand has adapted by shifting
to less dense and more peripheral neighborhoods. 

\subsection{Local Housing Markets and the May 2020 Urban Protests\label{subsec:Local-Housing-Markets}}

\begin{figure}
\caption{The U.S. Housing Market in 2020: George Floyd Protests and Urban Housing
Markets}

\begin{center}

\subfloat[The Spatial Extent of the May 2020 Protests\label{fig:Riots-in-2020-vs-Riots-in-1968}]{

\includegraphics[scale=0.2,trim={6cm 15cm 6cm 15cm}, clip]{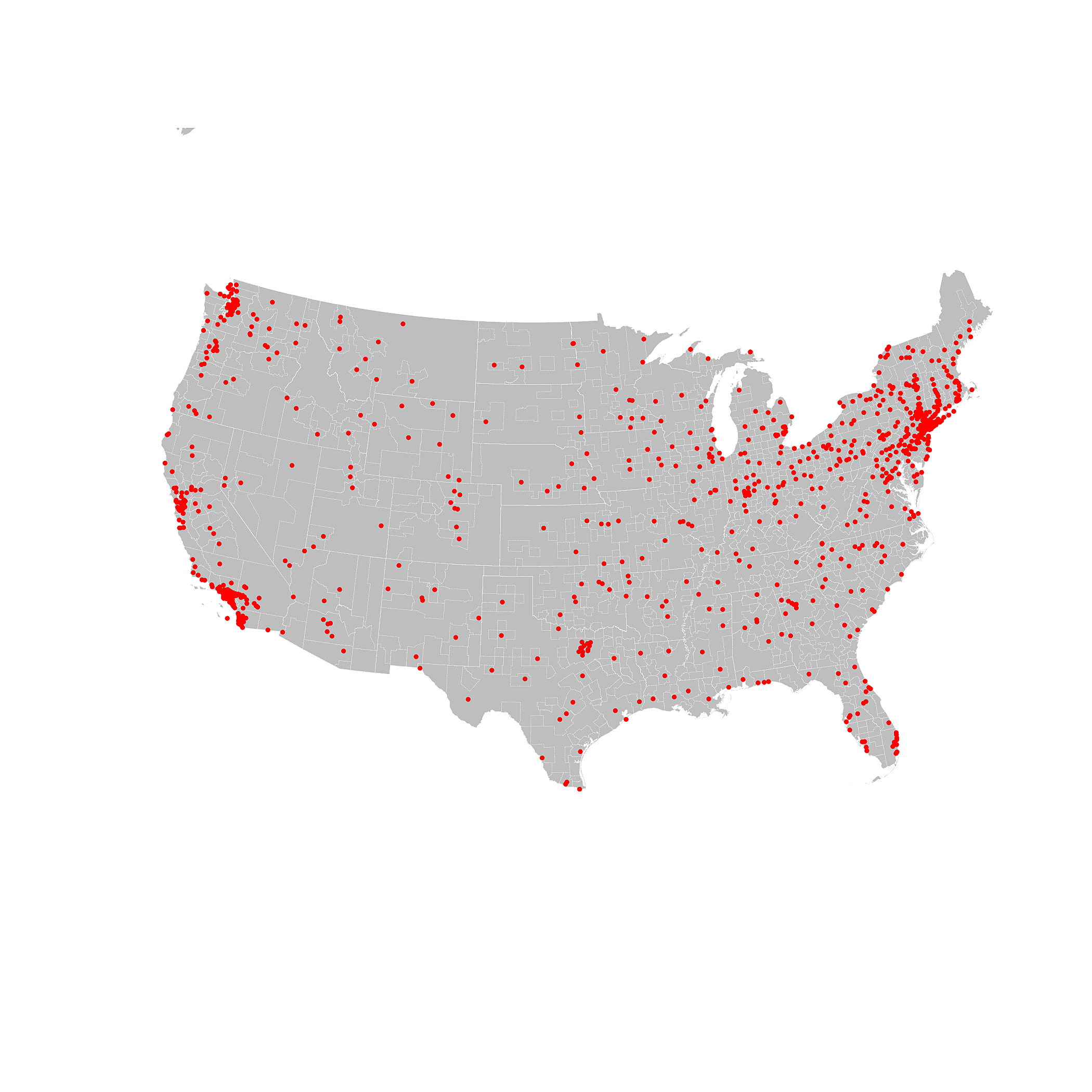}}

\subfloat[George Floyd Protests in Los Angeles\label{fig:George-Floyd-Protests-LA}]{

\includegraphics[scale=0.4]{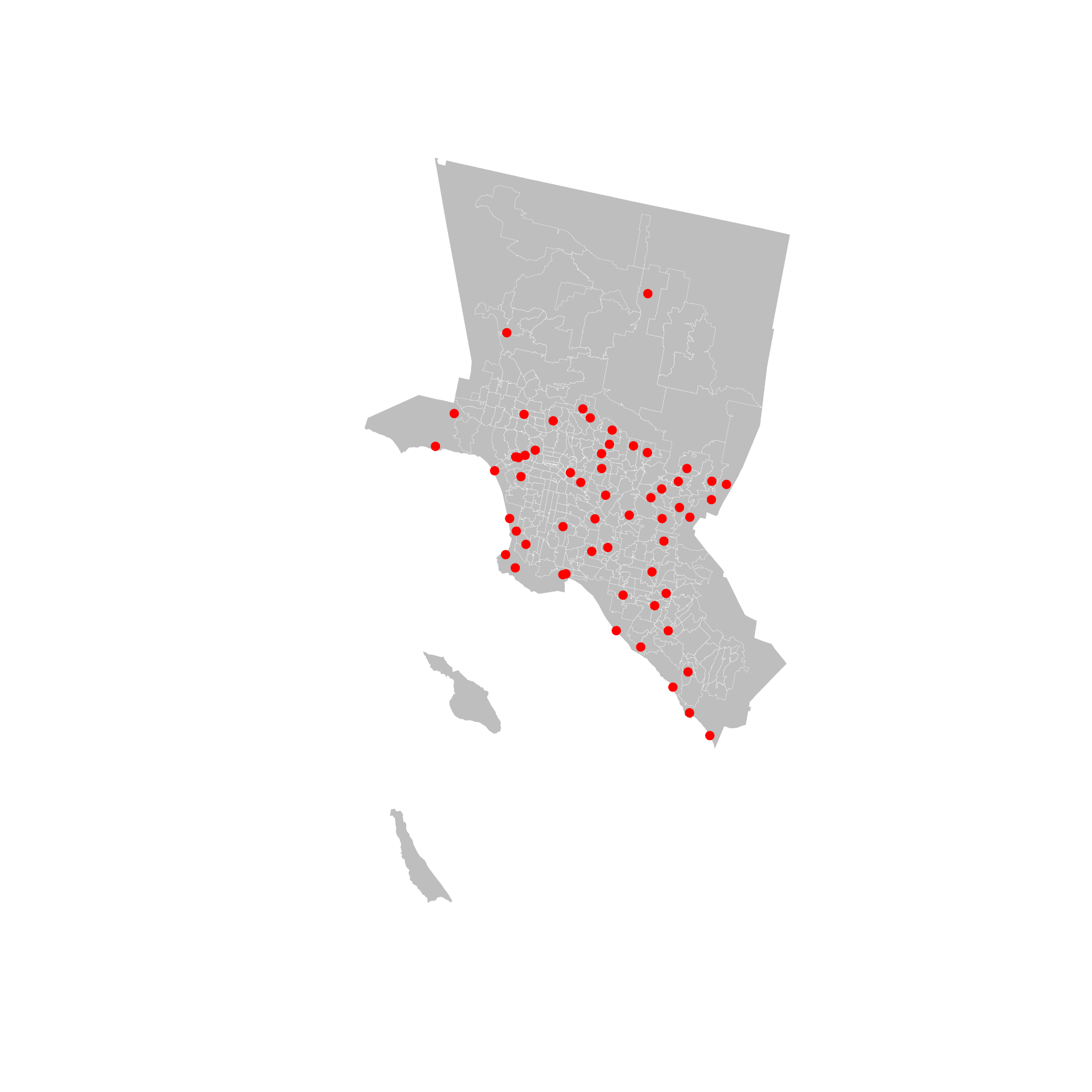}}\subfloat[Comparing Price Appreciation Across Neighborhoods in Los Angeles\label{fig:Comparing-Price-Appreciation}]{

\includegraphics[scale=0.4]{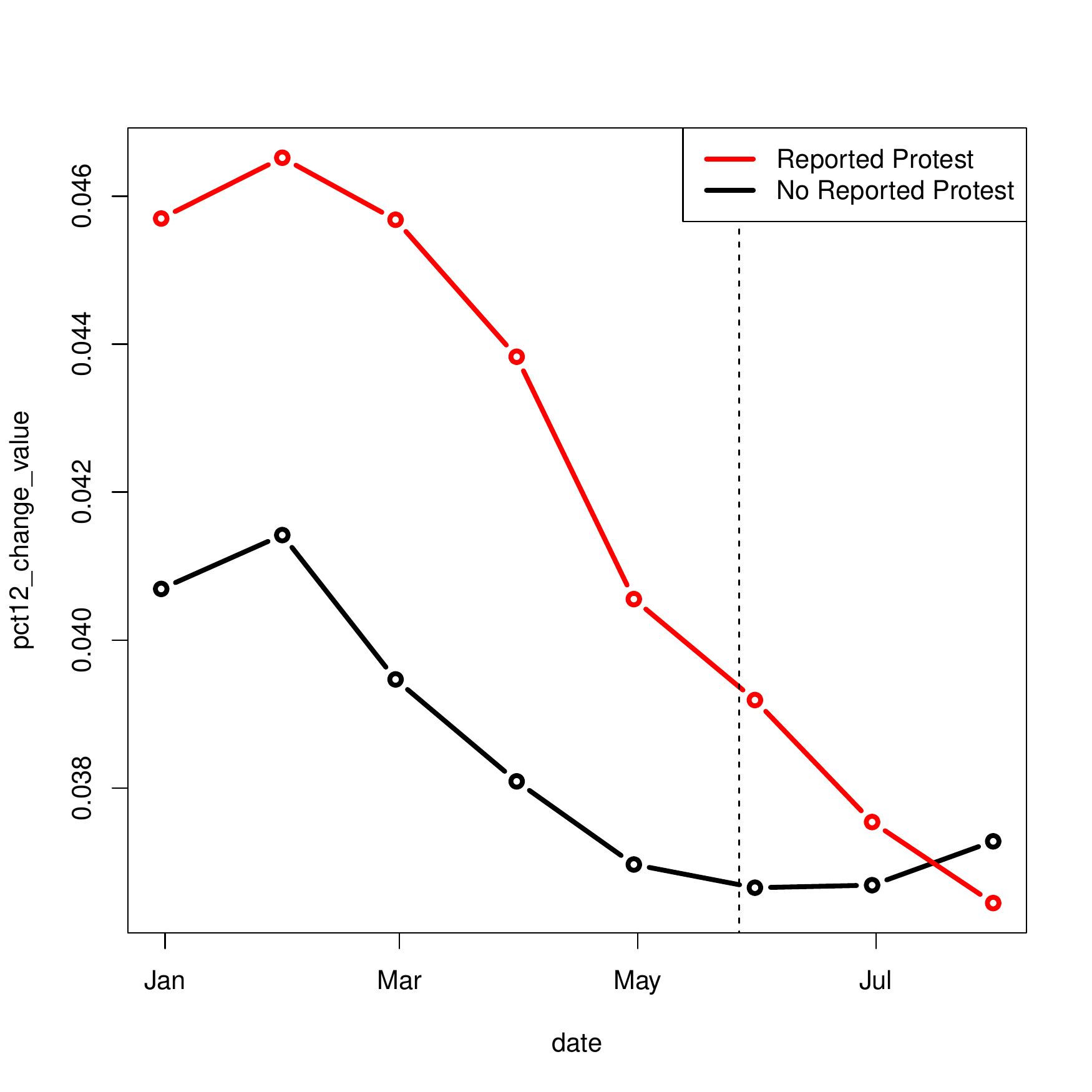}}

\end{center}

\emph{Source: Crowdsourced May 2020 George Floyd protest data through
the Wikimedia foundation. }
\end{figure}

The year 2020 saw a second series of shocks affecting urban housing
markets. Urban protests in response to alleged actions by the police
started in May 2020 and quickly spread to a substantial number of
U.S. metropolitan housing markets. Figure~\ref{fig:Riots-in-2020-vs-Riots-in-1968}
presents the geographic location of the May 2020 protests with more
than 100 participants according to the geocoded crowdsourcing of the
Wikimedia foundation.\footnote{Other potential sources of recent geocoded data include the \emph{Crowd
Counting Consortium}. Further literature may focus on \emph{Factiva}'s
news archive as an alternative source of information on protests.} The spatial extent of these protests exceeds those of the 1968 protests
as documented by Stanford University's Susan Olzak in her collection
of \emph{Ethnic Collective Action in Contemporary United States. }This
suggests that the 2020 urban protests may be the largest protests
in U.S. history. Whether protests lead to positive reforms that improve
the desirability of urban living; or whether protests lower the quality
of life in urban metros is an empirical question. 

\citeasnoun{collins2007economic} uses decennial Census data between
1950 and 1980 to describe the long-term impact of the 1960s riots
on property values. They suggest that riots led to a decline of property
values, and in particular to a decline in black-owned property values,
with no rebound in the 1970s. The perhaps most salient example of
such decline is the city of Detroit. In this context, \citeasnoun{glaeser2005urban}
argues that shocks may lead to a long decline in metropolitan areas
as the supply curve of housing is L-shaped: a decline in housing demand
may lead to a decline in house prices down to the marginal cost of
housing, leading to larger vacancy levels, attracting lower productivity
workers and lowering the benefits of agglomeration economies. 

Figures~\ref{fig:George-Floyd-Protests-LA} and \ref{fig:Comparing-Price-Appreciation}
present a correlational analysis of urban protests and house prices
in the metropolitan area of Los Angeles. Figures~\ref{fig:George-Floyd-Protests-LA}
presents evidence that George Floyd protests extended from the northern
neighborhood of San Fernando to the southern neighborhoods of Laguna
Niguel. Figure~\ref{fig:Comparing-Price-Appreciation} compares house
price appreciation in ZIP codes where a protest happened (red line)
to house price appreciation in ZIP codes where a protest did not occur
(black). While the hypothesis that the appreciation rates are parallel
cannot be rejected statistically prior to May 2020, the appreciation
rate declines and \emph{crosses }the appreciation rate of ZIP codes
where a protest did not occur. Hence, while on average across the
United States, house price increases suggest expectations of urban
resilience, there is local evidence of some expectations of decline
in specific neighborhoods affected by the urban protests. This may
be driven by the shift of demand towards neighborhoods less exposed
to risk.

The endogeneity of riots may cast doubt on the causal interpretation
of such event studies that rely on a pre-post analysis of the impact
of riots on urban growth and decline. \citeasnoun{dipasquale1998angeles}
finds support for a Beckerian mechanism in which protests are the
outcome of a comparison between the opportunity cost of time and the
potential cost of punishment, and consistent with evidence by \citeasnoun{esteban2011linking},
the paper finds that ethnic diversity matters. In the case of Los
Angeles in May 2020, evidence suggests significant differences in
the demographics of Zips with urban protests and without urban protests.
\begin{center}
{\footnotesize
\begin{tabular}{lcccccc}
\toprule
 & \multicolumn{2}{c}{\textbf{Mean}} \\
\cmidrule(lr){2-3}
 & \textbf{Protest Zips} & \textbf{Other Zips} & \textbf{Difference} & \textbf{S.E.} & \textbf{t} \\    \midrule
Frac. African American &  0.054 &  0.084 & $-$0.029 & (0.019) & $-$1.56 \\    
Frac. Hispanic &  0.372 &  0.406 & $-$0.034 & (0.041) & $-$0.82 \\
Frac. Asian &  0.219 &  0.186 &  $+$0.032 & (0.025) & $+$1.27 \\ 
Frac. Owner Occupied &  0.545 &  0.479 &  $+$0.066 & (0.033) & $+$1.98 \\
log(Median Household Income) & 11.257 & 11.116 &  $+$0.141 & (0.063) & $+$2.22 \\
Frac. Poverty &  0.124 &  0.154 & $-$0.030 & (0.016) & $-$1.82 \\   
Frac. No Health Coverage &  0.113 &  0.133 & $-$0.020 & (0.012) & $-$1.70 \\
\bottomrule
\end{tabular}}
\end{center}In particular, this table suggests that protests occurred in neighborhoods
that had significantly higher household income, lower shares of African
Americans and Hispanics, higher shares of owner-occupied housing,
lower poverty rates, and lower fractions of households with no health
coverage. In the future, longer time series combined with sound identification
strategies may allow for a causal analysis of the 2020 protests on
urban housing markets.

\section{Housing Markets in the Long Run:\protect \\
The Role of Shocks vs. Initial Conditions\label{sec:Looking-Forward}}

The previous section described the short-run response of U.S. housing
markets to the pandemic and the urban protests. The long-run prospect
is yet unknown. The past can nevertheless inform our perception of
future trends. This section describes the long-run evolution of metropolitan
areas between 1970 and 2010 using longitudinal census tract data.
It sheds light on the drivers of the rise and decline of cities. Are
cities that experience large short-run shocks rebounding or are the
typical impacts permanent shifts in population levels? Prior literature~\cite{gabaix1999zipf,ioannides2003zipf}
has described the relative stability of city size distributions, which
follow Zipf's law, where the log population is a linear relationship
to the log rank of the metropolitan area. Yet, within such distribution,
metro areas rise and fall. Understanding which observable characteristics
drive such rise and fall is the focus of the first section~\ref{subsec:Aggregate-City-Dynamics:}.
While metropolitan area rankings tend to be stable, the desirability
of specific neighborhoods within metropolitan areas changes more dramatically
over time. This is the focus of subsections~\ref{subsec:The-Ebb-of}
and~\ref{subsec:Case2_sf_earthquake}. We present two case studies:
the New York housing market in the aftermath of September 11; and
the dynamic of San Francisco's neighborhoods after the 1989 Loma Prieta
earthquake.

\subsection{Explaining Metropolitan Growth in the Long Run\label{subsec:Aggregate-City-Dynamics:}}

The relative ranking of metropolitan areas is stable over time: data
from the \emph{Neighborhood Change Database} suggests that the correlation
between a metropolitan area's population rank in 1970 and its rank
in 2010 is 0.8, implying that the best predictor of a city's future
is its past. Rankings are also stable in other dimensions than population:
\citeasnoun{kerr2020tech} shows that 1975-1980 annual patent count
is a strong predictor of 2013-2018 patent count. Yet, some metropolitan
areas experience rapid population shifts: the Dallas-Fort Worth-Arlington
went from being the 11th most populus metro to the number 4 rank.
The Atlanta metropolitan area went from the 19th to the 9th rank,
joining the 10 largest metro areas. In contrast, Pittsburgh went down
13 notches, from the 9th most populous metro to the 22nd most populous,
as the steel industry declined. Two of the largest relative growth
levels were observed in Las Vegas, going from the 102nd to the 31st
largest; and the Austin\textendash Round Rock metropolitan area, jumping
58 spots to the 35th rank.

The largest relative decline is that of Johnstown, Pennsylvania going
from the 150th to the 249th spot, with a 50.6\% decline in population;
this metropolitan area experienced three major floods, the most recent
in 1977. This major flood could be a candidate for a causal driver
of the city's decline. Another competing explanation for this decline
is Johnstown's \emph{specialization} in the steel industry, with steel
mill plants in the heart of its downtown.

Hence, for Johnstown as for other metropolitan areas, a key question
is whether shocks (here floods) or fundamentals (here industrial composition)
explain their rise and fall? We use data from a range of sources to
estimate the correlation between urban growth and (i)~natural disasters,
(ii)~urban protests, (iii)~industrial composition, (iv)~education
levels, (v)~urban segregation, and (vi)~housing supply elasticity.
Each of these hypothesis has received support in the literature. The
analysis of this chapter is not comprehensive, yet provides an overview
of the potential drivers of urban growth and decline. 

\subsubsection*{The ``Shocks'' Hypothesis}
\begin{itemize}
\item \emph{Natural Disasters}
\end{itemize}
Natural disasters may cause either temporary or permanent shifts in
population levels. We use data from NOAA's significant storm events,
which provides damages and fatalities at the county level since January
1950. We count the number of billion dollar storms for each county
in the 1970-2010 period. The metropolitan area with the largest number
of such storms is the New Orleans\textendash Metairie, LA CBSA, with
12 billion dollar storms. Then comes the Gulfport-Biloxi-Pascagoula,
MS CBSA, with 6 such storms, and the Houston-The Woodlands-Sugar Land,
TX CBSA, with also 6 such storms. The New Orleans CBSA is also the
metropolitan area with the largest amount of billion dollar damages.
We consider three variables explaining metropolitan growth: the number
of events, the total property damages, and whether there was any event. 
\begin{itemize}
\item \emph{Urban Protests}
\end{itemize}
We test the urban protest hypothesis using data collected by Susan
Olzak on Ethnic Collective Action in the United States~\citeasnoun{olzak1995ethnic}.
The list of events was compiled from the New York Times Index and
from microfilms of New York Times articles. We focus on protests occuring
between 1970 and the last date of the file, 1992. The data reports
the number of protestors, the involvement of police, damage to property,
the presence of non residents, and other features, for each metropolitan
area. We match the now deprecated Standard Metropolitan Area (SMSA)
geographies to the 2010 Core Based Statistical Area, which is the
most recent definition of metropolitan boundaries.

\subsubsection*{The ``Fundamentals'' Hypothesis}

We compare the impact of shocks to the impact of the following fundamentals:
education, industrial composition, segregation, and housing supply
elasticity. In each case, we describe the associated literature and
the data used. 
\begin{itemize}
\item \emph{Industrial Composition}
\end{itemize}
Initial industrial composition may matter for long term metropolitan
growth through a number of channels. First, specialization in industries
with strong global demand for their products may lead to a greater
demand for labor in the metropolitan area. This is the intuition of
\citeasnoun{bartik1991benefits} and \citeasnoun{jean1992regional}.\footnote{For a discussion of this empirical approach, see \citeasnoun{goldsmith2018bartik}.}
Second, the diversity of industries initially present in a metropolitan
area may foster the growth of a variety of industries that depend
on an economic fabric of different suppliers and different customers.
This is the industrial diversification hypothesis, perhaps most saliently
popularized by Jane Jacobs in the Death and Life of Great American
Cities: ``\emph{Typically {[}small manufacturers{]} must draw on
many and varied supplies and skills outside themselves, they must
serve a narrow market at the point where a market exists, and they
must be sensitive to quick changes in this market. Without cities
they would simply not exist. {[}...{]} City diversity itself permits
and stimulates more diversity}.'' (Chapter 7, The Generators of Diversity).

We estimate the correlation between industrial composition (either
specialization or diversification) using the earliest wave of publicly
available data from the County Business Patterns. These data provide
establishment numbers for each Standard Industrial Classification
(SIC) 2-digit code. We aggregate such county level data to the boundaries
of 2010 Core Based Statistical Areas, the same boundaries as those
of the Neighborhood Change Database \textendash{} this allows for
measuring the growth of metropolitan areas. To test the specialization
hypothesis, we use 2-digit SIC codes, leading to the following categories:
Agriculture, Fishing and Forestry; Metal, Mining and Oil; Construction;
Manufacturing; Transportation and Utilities; Finance, Insurance, and
Real Estate; Non Classifiable; Retail. The measure of industrial specialization
is the Herfindahl index (HHI), which is equal to the sum of the squares
of the 2-digit SIC industry establishment shares:
\begin{equation}
HHI_{m}=\sum_{k}\left(Share_{k}\right)^{2},\label{eq:hhi}
\end{equation}
where $m$ is the metropolitan area, $k$ is the 2-digit SIC code,
and $Share_{k}$ is the proportion of establishments in industry $k$.
We use the share of establishments as this variable is well filled
in the US Census Bureau's County Business Patterns. Given the large
asymmetry and the fat tails of the $HHI$ measure, results of the
linear regression are more robust when regressing on four indicator
variables for the four quantiles of HHI, from least specialized (Q1),
to most specialized (Q4).
\begin{itemize}
\item \emph{Education}
\end{itemize}
In \citeasnoun{moretti2012new}, the author describes the diverging
paths of Menlo Park and Visalia, CA, and suggests that Menlo Park
experienced significantly stronger growth thanks to its higher share
of educated residents. In the \emph{Rise of the Skilled City}, \citeasnoun{glaeser2003rise}
describes the higher growth of more educated cities, even after controlling
for a range of covariates. This chapter's measure of education is
the fraction of college graduates in 1970, according to the 1970 Census
Count 4Pa, provided by the National Historical Geographic Information
System at the University of Minnesota. 
\begin{itemize}
\item \emph{Segregation and Inequality}
\end{itemize}
Our third measure of metropolitan area fundamentals is urban segregation.
A number of papers suggest that urban racial segregation affects welfare.
\citeasnoun{li2013residential} argues that urban segregation has
effects on metropolitan economic growth beyond its effects on minorities
and poor residents. Thus urban segregation may be a concern for both
distributional and efficiency reasons. \citeasnoun{card2007racial}
suggests that neighborhood segregation has a consistently negative
impact on the SAT scores of black students. \citeasnoun{watson2006metropolitan}
describes the negative correlation between income segregation and
metropolitan population growth.

We build a measure of Black-{}-White urban segregation in 1970, at
the beginning of our time period. The dissimilarity index measures
the difference between the distribution of black residents across
neighborhoods and the distribution of white residents across the same
set of neighborhoods. We use 1970 census tract demographics. The dissimilarity
index is a popular measure of segregation, notably developed in \citeasnoun{duncan1955methodological}
and used in \citeasnoun{cutler1999rise}. The dissimilarity measure
used in this paper is:
\begin{equation}
D_{m}=\frac{1}{2}\sum_{j}\left|\frac{w_{m,j}}{w_{m}}-\frac{b_{m,j}}{b_{m}}\right|,\label{eq:dissimilarity}
\end{equation}
where $m$ is one of the 306 metropolitan areas, $j$ indexes neighborhoods,
$w_{m,j}$ (resp. $b_{m,j}$) is the number of white (resp. black)
residents in neighborhood $j$, $w_{m}$ (resp. $b_{m}$) the number
of white (resp. black) residents in metropolitan area $m$. Results
using other pairs of races and ethnicities are available from the
author. Notable examples of segregated metropolitan areas include
the Chicago-Naperville-Elgin, IL-IN-WI metropolitan area ($0.90$),
Oklahoma City, OK ($0.89$), Los Angeles-Long Beach-Anaheim, CA ($0.89$),
and Detroit-Warren-Dearborn ($0.88$). Alternative segregation indices
such as the exposure or the normalized exposure indices~\cite{cutler1999rise,ouazad2016credit}
provide different rankings, yet these three indices are strongly correlated. 
\begin{itemize}
\item \emph{Housing Supply Elasticity}
\end{itemize}
Our final hypothesis is that constraints on housing supply, stemming
either from geographic or regulatory constraints, are a barrier to
the development of metropolitan areas; they indeed constrain the growth
of the housing stock~\cite{mayer2000residential,glaeser2006urban,saks2008job},
and make housing more expensive for productive workers whose productivity
gains are transferred to the owners of land.

There is a variety of available housing supply elasticity measures,
starting with~\citeasnoun{saiz2010geographic}. We use recent metro-level
elasticity measures from \citeasnoun{gorback2020global}, yet using
\possessivecite{saiz2010geographic} measures does not affect the
regression estimates. We control for an indicator variable for a missing
elasticity measure, as housing supply elasticity is typically not
available for smallest metropolitan areas.
\begin{itemize}
\item \emph{Other possible fundamentals}
\end{itemize}
Other fundamentals could be included in a further analysis: innovations
measured by the number of patents per capita~\cite{kerr2020tech},
market access and transportation costs~\cite{redding2010empirics},
public transportation infrastructure~\cite{kahn2007gentrification,gonzalez2018subways},
the proximity to deepwater ports~\cite{brooks2018local}, the flow
of credit due to the structure of the banking sector in the metropolitan
area~\cite{clarke2004geographic,ouazad2016credit}, and other fundamentals. 

\subsubsection*{Estimation Results: Shocks and Fundamentals}

Results of the analysis are presented in Table~\ref{tab:Within-City-Adaptation:-Populati}.
The first columns present the covariates separately (education, industrial
composition, segregation, elasticity, shocks), and the last columns
performs the regression with all previous covariates simultaneously.
In all 11 regressions the dependent variable is the change in the
metropolitan area population rank between 1970 and 2010. A first notable
fact is the strong correlation of black-white urban segregation, education,
and industrial specialization, with a metropolitan area's relative
growth. More segregated areas grow less than other, more integrated
areas. Metropolitan areas with larger shares of college-educated residents
grow significantly more. Areas with less diverse industrial composition
(an HHI in the 4th quartile) tend to grow significantly less \textendash{}
consistent with Jane Jacobs' hypothesis. Regressions indicate that
it is the concentration in one or a few industries that predicts urban
decline rather than the specialization in manufacturing. 

Notably, none of the shocks \textendash{} urban protests and storms
\textendash{} have a statistically significant impact at 5\%. There
is no significance whether one looks at the number of riots, whether
there is any riot, the dollar amount of property damages due to storms,
the number of storms, or whether there is any storm. In some cases
the sign is as expected: a larger number of riots with damages to
property has a negative impact on a metropolitan area's population
growth; yet the impacts are not significant. 

The last column includes all of the previous covariates simultaneously.
Interestingly, both urban segregation and industrial specialization
remain strongly significant (at 1\%), again consistent with the central
tenets of Jane Jacobs' the \emph{Death and Life of Great American
Cities}. Shocks remain non significant. Perhaps notable is the significance
of the housing supply elasticity measure: when controlling for other
fundamentals, metropolitan areas with higher housing supply elasticities
experience significantly higher growth (significant at 5\%).

\subsection{The Resilience of the New York City Housing Market After September
11\label{subsec:The-Ebb-of}}

While city rankings by population size are stable, the ranking of
neighborhoods tends to fluctuate substantially over time. Evidence
from the Neighborhood Change Database suggests that the correlation
between a tract's ranking in 1970 and the same population ranking
in 2010 is only $0.2$. This suggests that cities may be resilient
when urban residents adapt their location and housing consumption
by using the variety of amenities, housing stocks, and access to jobs
to respond to shocks.

September 11 2001 presents a case study for the impact of a terrorist
event on the desirability of living in dense urban spaces. The event
had dramatic consequences on the welfare of central New York City
residents: \citeasnoun{galea2002psychological} suggests that adults
experienced symptoms consistent with post traumatic stress disorder
(PTSD), with a prevalence of PTSD up to 20\% for those living south
of Canal Street near the World Trade Center. In a set of respondents
with an oversampling of children new the World Trade Center, \citeasnoun{hoven2005psychopathology}
finds that 29\% of children experienced anxiety disorders. 

This may have impacts on the New York housing market. In Israël, \citeasnoun{elster2017rockets}
using hedonic and repeat sales approaches to show that attacks led
to a 6 to 7\% decline in house prices and rents. They also find that
these effects are perisistent beyond the 2000-{}-2012 period, and
suggest this is consistent with a perception of a continued threat.
\citeasnoun{bram2004has} suggests that the September 11 events caused
a sharp contraction of business activity. In the long run, \citeasnoun{eisinger2004american}
claims that ``few lasting effects on city life are evident,'' and
suggests that city dynamics are affected by long-term forces rather
than even very significant short term ones.

We provide quantitative neighborhood-level evidence of the dynamics
of housing markets during and in the aftermath of September 11 using
5-digit ZIP code price data since 1996. We are thus able to estimate
pre-existing trends, the impact of the events during the September
to December 2001 period, and during the post 2001 period. We can also
test whether these events affected the desirability of central city
living in New York. 

Evidence suggests a strong rebound of price growth in the October
to December 2002 period compared to the October to December 2001 period.
Monthly year-on-year price appreciation for the New York MSA as a
whole and for the central New York City ZIP codes suggests that prices
increased significantly a year after September 11 2001. There is no
immediate discernible negative impact of September 11 on price appreciation
in the New York MSA as a whole, suggesting that even shocks that a
have strong negative impact on residents' welfare have not been capitalized
into long run house prices. 

\begin{figure}

\caption{Within-City Adaptation to Shocks: Short-Run Suburbanization in NYC
In September-December 2001\label{fig:Price-Appreciation-by-Distance-to-the-Center-Sep-11}}

\emph{These four graphs present the average price appreciation (using
the ZHVI index) for bins of neighborhoods ordered by their distance
to the Central Business District of the New York metropolitan area.
Figures (b) and (c) suggest that the relationship changed sign, before
going back to the average negative gradient observed prior to September
11. }

\bigskip{}

\subfloat[Before 2001: Higher Price Increase in the CBD\label{fig:Before-September-2001:}]{

\includegraphics[scale=0.45]{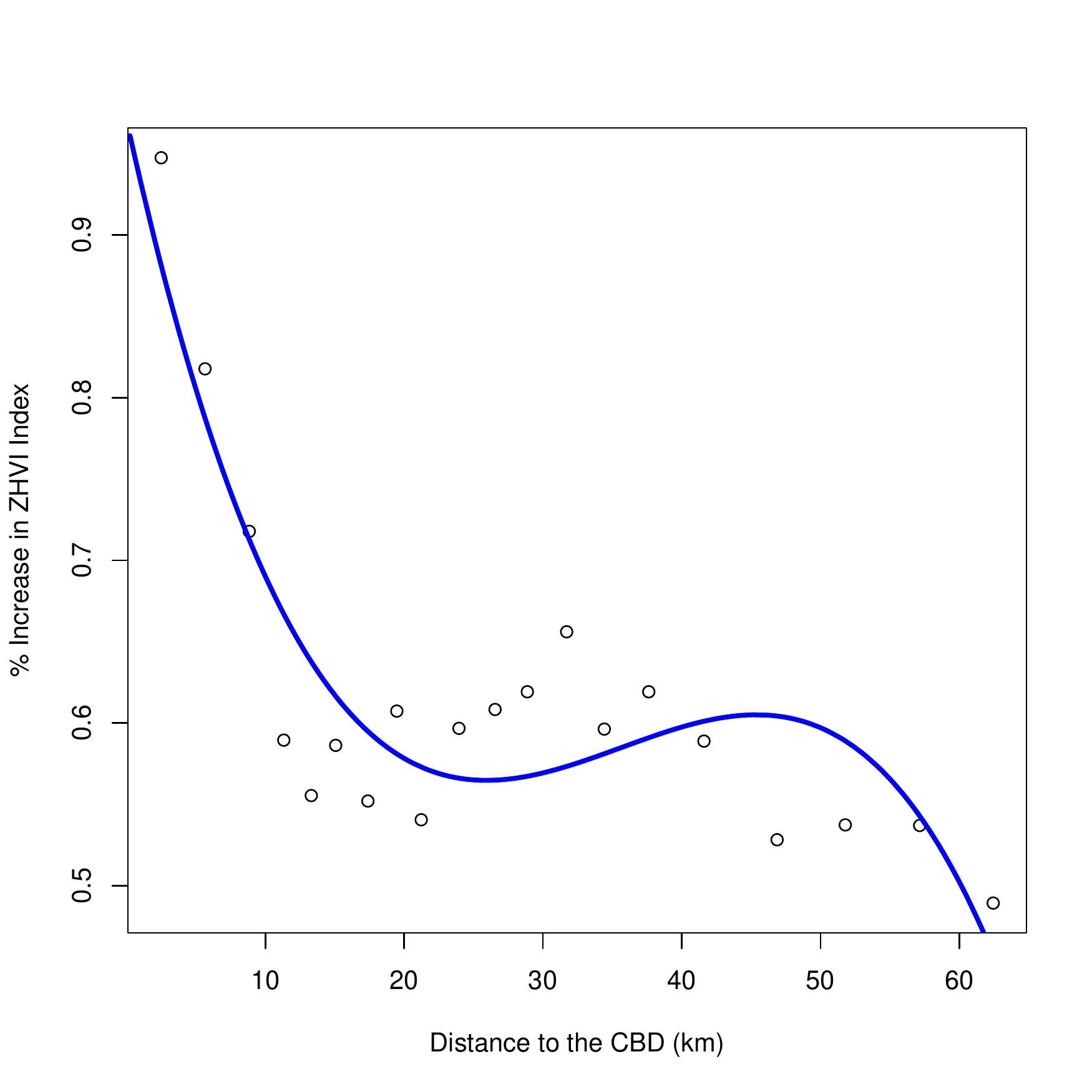}}\subfloat[2002 to 2020: Higher Price Appreciation in the CBD]{

\includegraphics[scale=0.45]{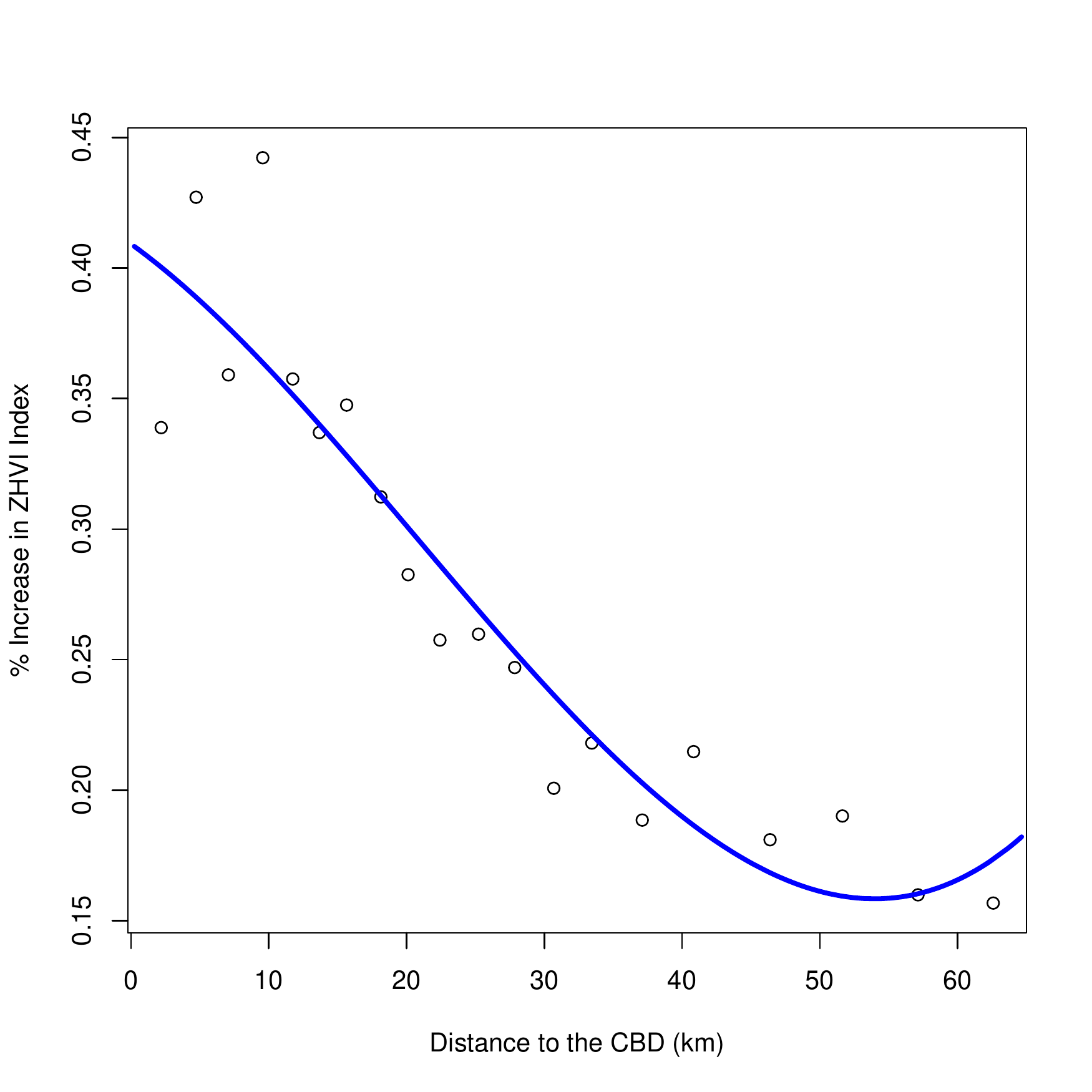}}

\bigskip{}

\subfloat[September 2001: Appreciation in the Periphery]{

\includegraphics[scale=0.45]{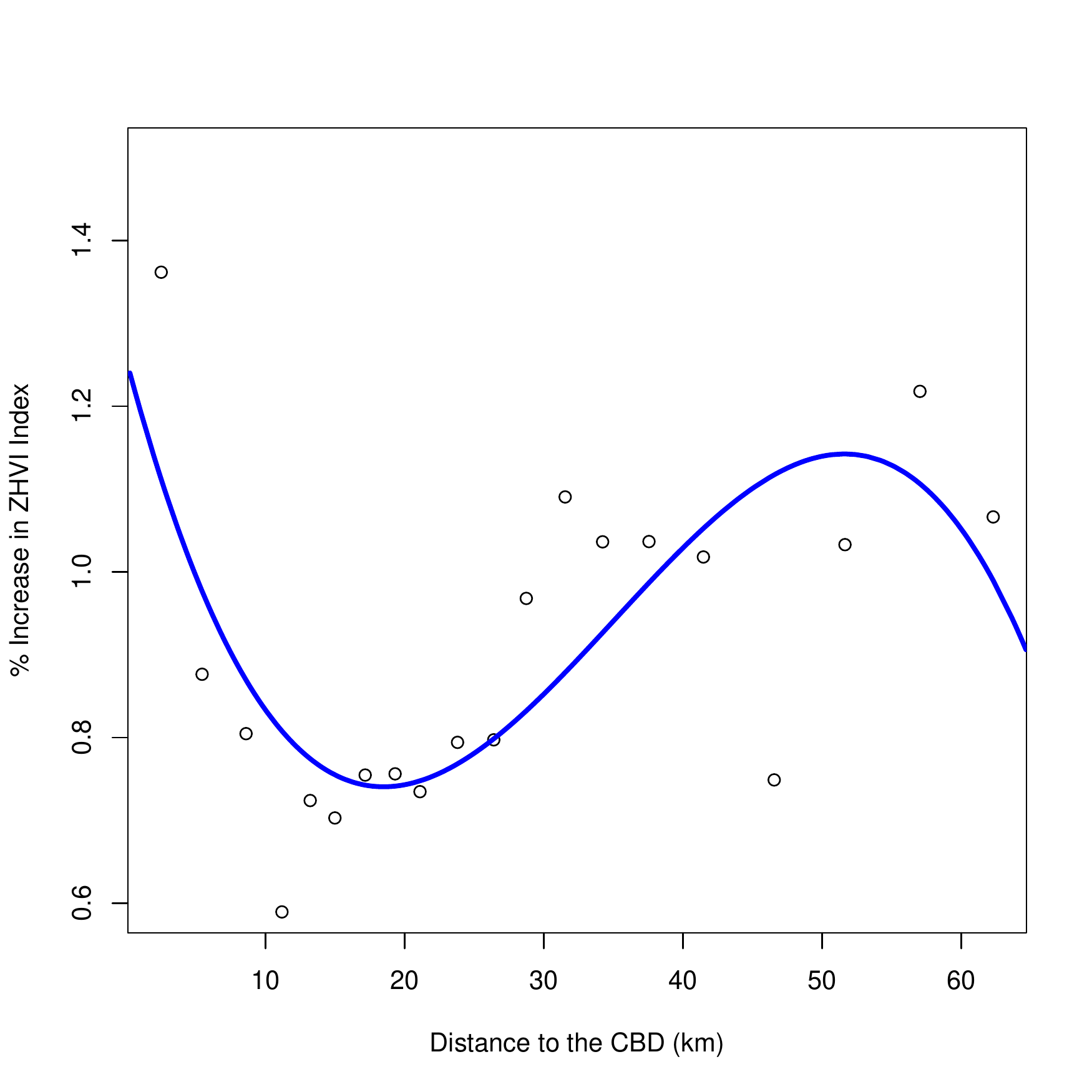}}\subfloat[October 2001: Appreciation in the Periphery]{

\includegraphics[scale=0.45]{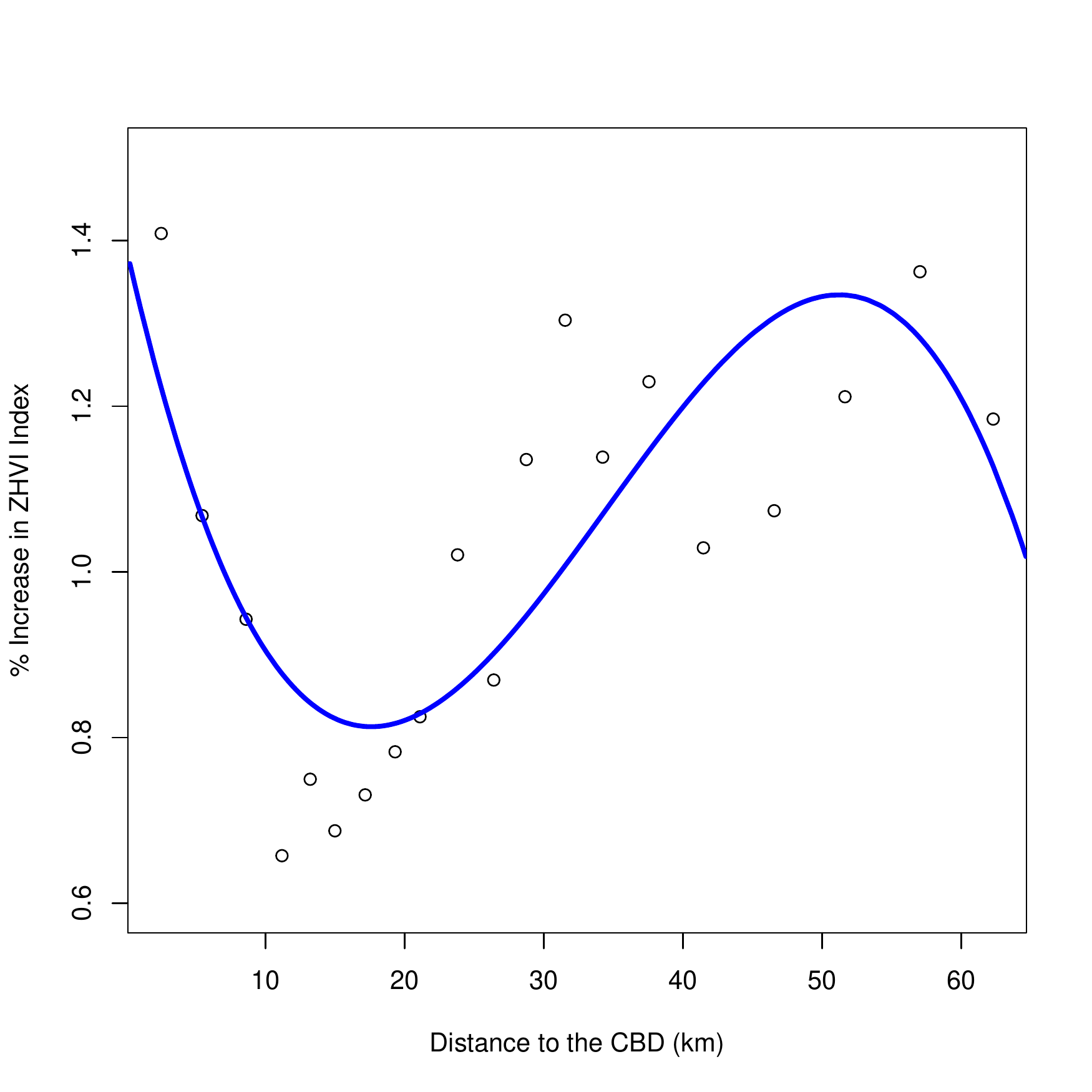}}

\emph{Zip-level ZHVI index from Zillow. Appreciation is month to month
in this graph.}
\end{figure}

Yet, Figure~\ref{fig:Price-Appreciation-by-Distance-to-the-Center-Sep-11}
does present evidence that September 11 temporarily shifted demand
away from central New York City and to the suburbs. Panel (a) shows
that price appreciation is up to twice stronger in ZIP codes close
to the central business district than for neighborhoods in the 10
to 60 kilometer range (6 to 37 miles) from the central business district.
This relationship is almost flipped in September and October 2001,
wher price appreciation is larger in ZIP codes farther away from the
Central Business District (CBD) than close to it. Yet, panel (d) suggests
this is only a temporary phenomenon, as price appreciation is again
decreasing with the distance to the CBD between 2002 and 2020. This
evidence suggests that while September 2001 did affect the demand
for central New York residential housing, these effects did not last
beyond 2001, at least in terms of price appreciation for residential
units in the densest parts of New York. 

The dramatic shock of September 2001 also affected the demand for
central city residential housing in other cities. \citeasnoun{abadie2008terrorism}
suggests that 9/11 increased Chicago's residents perception of the
probability of terrorist attacks. They show that vacancy rates increased
in the vicinity of the Sears Tower, the Aon Center, and the Hancock
Center. This section's result do not however provide evidence of long
run impacts of these events on residential housing markets.

\subsection{Rebuilding San Francisco After the 1989 Loma Prieta Earthquake\label{subsec:Case2_sf_earthquake}}

The Loma Prieta earthquake was an earthquake of magnitude 6.9 on the
Richter scale that shook the San Francisco Bay area on October 17,
1989. According to the California Department of Conservation, it caused
63 fatalities, 3,737 injuries, and 6 billion dollars in property damage.
Its epicenter was only 32.5 miles from Cupertino and 48 miles from
Menlo Park, both of which were and still are, major centers of technological
innovation.

A study published in the years following the earthquake \cite{murdoch1993impact}
analyzed the dynamic of house prices in six counties that were affected.
The study used all residential home sales between January 1988 and
November 1990. Results controlling for a substantial range of covariates
suggested that the disaster caused an overall decline in property
values as well as a gradient between house prices and measures of
earthquake risk such as soil type and seismic zone designation. Yet,
a key question is whether these price declines persisted and whether
local amenities were affected in the long run.

In this last section, we perform an analysis of the long-run impact
of the earthquake on neighborhood-level population flows using data
from the California Conservation Department\footnote{CGS Information Warehouse: Regulatory Maps.}
on earthquake risk, and data from the Neighborhood Change Database.
In a first step, we estimate the liquefaction risk for each block
of the San Jose-San Francisco-Oakland Combined Statistical area. According
to the Geological Survey, liquefaction takes place ``when loosely
packed, water-logged sediments at or near the ground surface lose
their strength in response to strong ground shaking.''\footnote{``What is liquefaction?'', Natural Hazards, U.S. Geological Survey.}
Liquefaction risk is a predictor of damage to structures~\cite{cubrinovski2011soil,towhata2016qualification}
as the nature of the soil leads to greater impacts on land at a given
earthquake magnitude.

In a second step, we matched such block-level liquefaction data with
the Neighborhood Change Database's tract level population levels.
We compute the share of a tract's area that is in the liquefaction
area. Prices are harder to analyze over such a long period nevertheless
population level are an indicator of the immediate impact of the earthquake
on living conditions, and long-term population changes are an indicator
of the quality of neighborhood amenities. \citeasnoun{owens2020rethinking}
argues that neighborhood population levels can decline below a threshold
that yields large amounts of vacancies.

Table~\ref{tab:Within-City-Adaptation:-Populati} indeed suggests
that population declined significantly in the immediate aftermath
of the earthquake. Census data was collected in 1990, only a few months
after the earthquake that shook the metropolitan area in October 1989.
The first column of the upper panel of the table suggests that population
declined 12\% between 1980 and 1990 in tracts that are entirely in
the liquefaction area. This is significant at 1\%. The first column
of the lower panel provides the regression where the dependent variable
is the tract's population rank. A tract within the liquefaction area
lost 35.9 ranks on average in 1990. Columns 2 and 3 nevertheless suggest
that the effect of the earthquake is relatively short-lived: tracts
in the liquefaction area experience no different population growth
in the two decades following the devastating earthquake. There is
no straightforward evidence that the earthquake is a major long-term
driver of population dynamics.

\begin{figure}
\caption{Within-City Adaptation to Shocks: The SF Bay Area After the 1989 Loma
Prieta Earthquake\label{fig:SF1989-Earthquake}}

\emph{Table~\ref{tab:Within-City-Adaptation:-Populati} showed that
liquefaction areas, while losing population compared to the rest of
the metropolitan area between 1980 and 1990, display no significantly
different population growth trend in the next decades (90s and 2000s).
These two maps show that indeed, population growth in 1990\textendash 2000
in Mountain View is not discontinuous at the border of the liquefaction
area.}

\begin{center}

\subfloat[Liquefaction Areas]{

\includegraphics[scale=0.8,trim={2cm 0cm 2cm 0cm},clip]{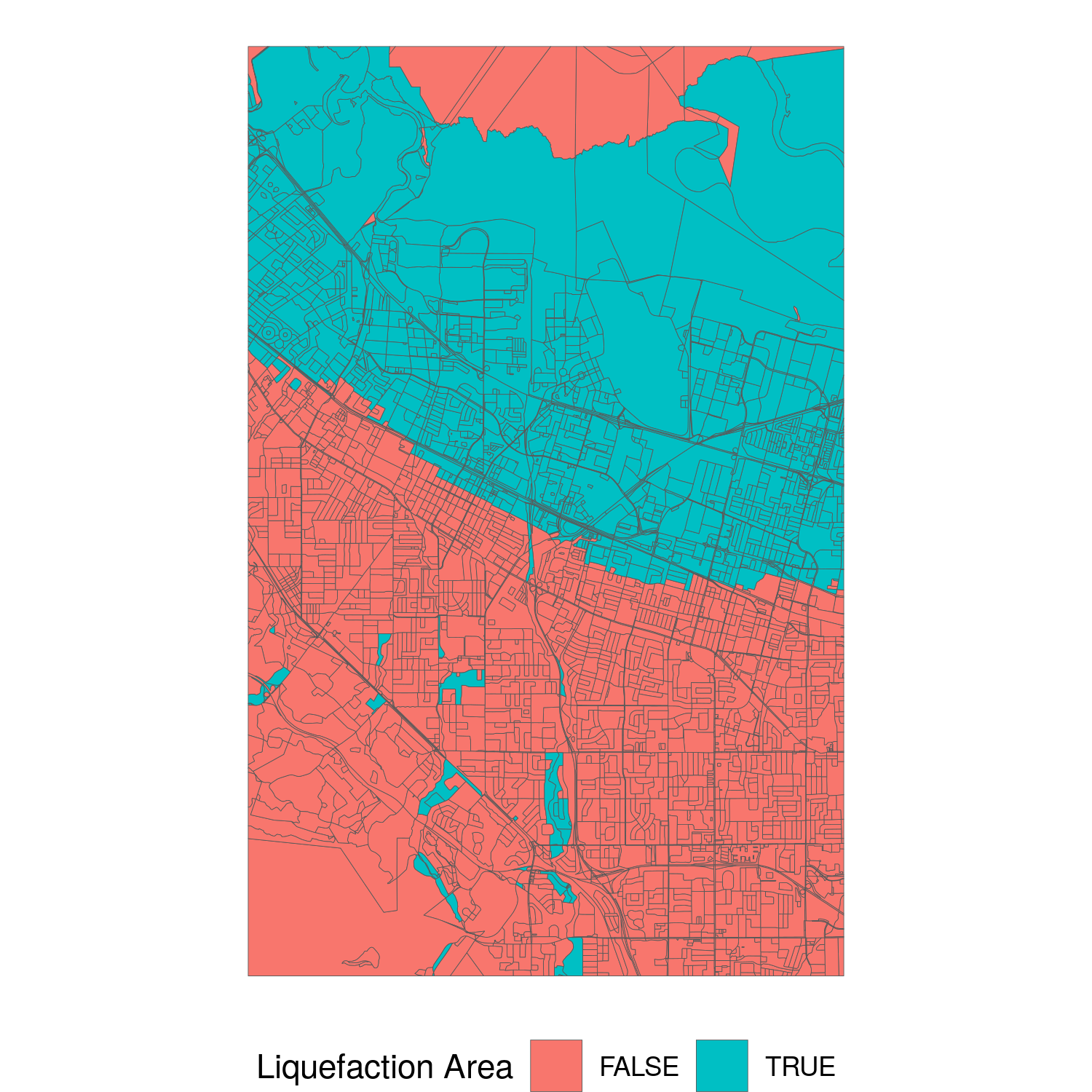}}\subfloat[Population Changes 1990-2000]{\includegraphics[scale=0.8,trim={2cm 0cm 2cm 0cm},clip]{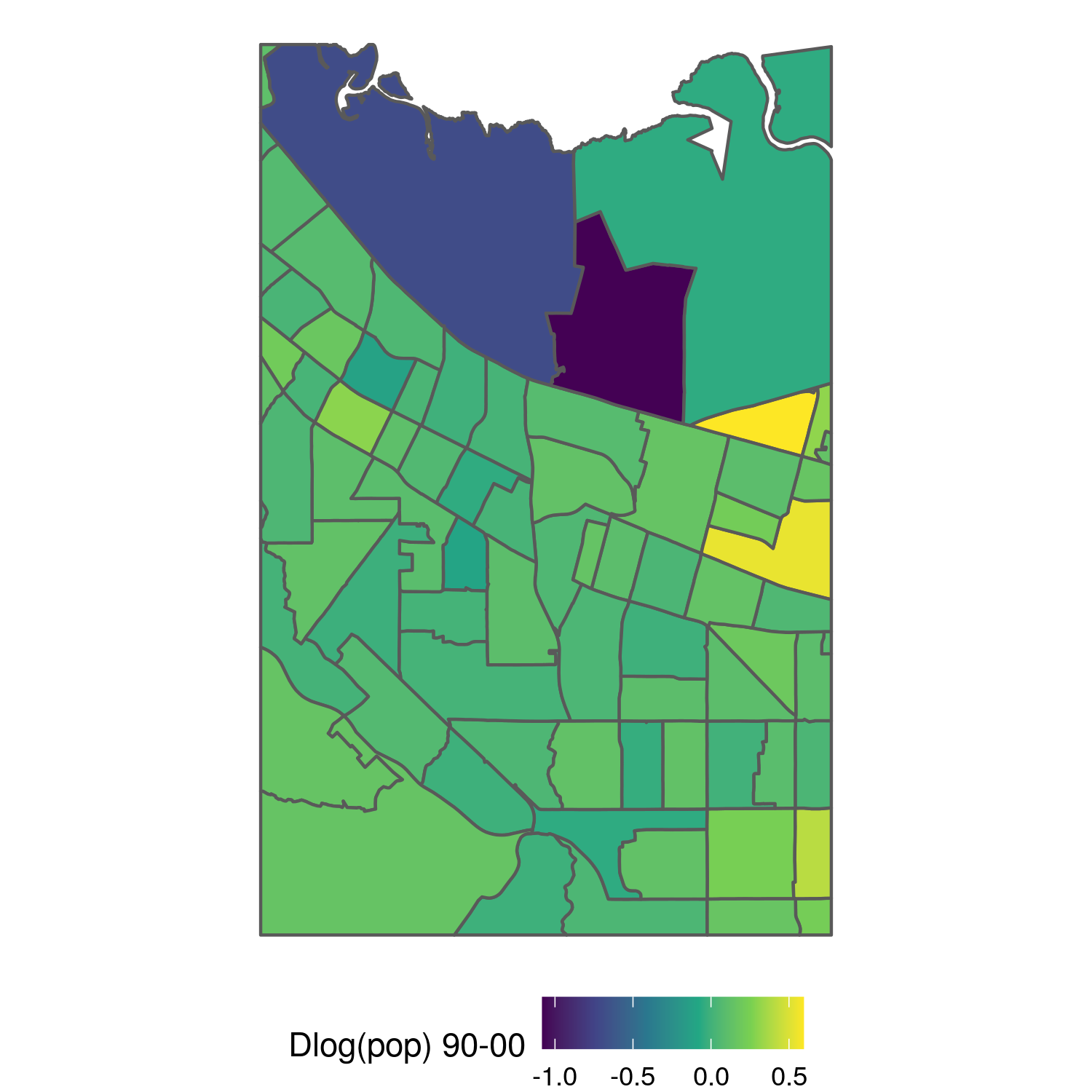}}

\end{center}

\emph{Source: California Department of Conservation's regulatory liquefaction
maps (left), matched to 2010 Census blocks. Geolytics Neighborhood
Change Database 1990-2000 at the tract level (right). }
\end{figure}

This is also clear in Figure~\ref{fig:SF1989-Earthquake}, which
focuses on Mountain View. While a substantial share of Mountain View
is in the liquefaction area, including the headquarters of Google
at 1600 Amphitheatre Parkway, there is no discernible impact of the
liquefaction area on population dynamics. In other words, a regression
discontinuity design at the boundary of such area would likely yield
no significant impact. This suggests that the 1989 Loma Prieta earthquake,
with damages estimated to 6 billion dollars~\cite{stover1993seismicity},
had only a minor impact on the San Francisco Bay Area's long term
population trend. 

\section{Conclusion\label{sec:Conclusion}}

The total magnitude and the length of both the Covid-19 pandemic and
the urban protests are, at the time of writing this chapter, yet unknown.
The past can nevertheless provide a sliver of hope for the future.
The evidence and the literature presented in this chapter suggest
that, over the span of four decades, metropolitan areas are remarkably
resilient to shocks \textendash{} fundamentals rather than short-run
shocks drive long-run population trends. Such resilience of urban
housing markets suggests that the benefits of agglomeration play a
key role in residents' welfare; sharing, matching, and learning are
key motives that explain the desirability of urban living. These benefits
have, over the long run, arguably been greater than the negative externalities
of agglomeration. High levels of education, a diversified industrial
composition, and racially integrated neighborhoods are keys to the
resilience of metropolitan areas. 

\bibliographystyle{agsm}
\bibliography{resilience}

\clearpage{}

\pagebreak{}

\clearpage{}

\pagebreak{}

\clearpage{}

\pagebreak{}

\clearpage{}

\pagebreak{}

\clearpage{}

\pagebreak{}

\clearpage{}

\pagebreak{}

\clearpage{}\pagebreak{}

\clearpage{}\pagebreak{}

\clearpage{}\pagebreak{}

\begin{table}
\caption{Confirmed Covid-19 Cases Per Capita and County Demographics\label{tab:regression_drivers_covid19_infections}}

\emph{This table correlates county Covid-19 cases per capita with
population density and Census demographics. Regressions include a
state fixed effect.}

\begin{center}

{\scriptsize 
\begin{tabular}{l c c c c c c c c}
  \toprule
  & \multicolumn{8}{c}{Dependent Variable: Confirmed Covid-19 Cases Per Capita} \\
 & (1) & (2) & (3) & (4) & (5) & (6) & (7) & (8) \\
\midrule
log(Density)                   & $0.13^{***}$ &               & $0.04^{*}$    & $0.03^{*}$    & $0.04^{**}$   & $0.05^{**}$   & $0.07^{***}$  & $0.07^{***}$  \\
                               & $(0.01)$     &               & $(0.02)$      & $(0.02)$      & $(0.02)$      & $(0.02)$      & $(0.02)$      & $(0.02)$      \\
Median Age                     &              & $-0.06^{***}$ & $-0.03^{***}$ & $-0.02^{***}$ & $-0.02^{***}$ & $-0.02^{***}$ & $-0.02^{***}$ & $-0.02^{***}$ \\
                               &              & $(0.00)$      & $(0.00)$      & $(0.00)$      & $(0.00)$      & $(0.00)$      & $(0.00)$      & $(0.00)$      \\
log(Median household income) &              &               & $-0.15$       & $0.45^{**}$   & $0.53^{***}$  & $0.50^{**}$   & $0.64^{***}$  & $0.64^{***}$  \\
                               &              &               & $(0.09)$      & $(0.15)$      & $(0.15)$      & $(0.15)$      & $(0.16)$      & $(0.16)$      \\
Frac. Black                    &              &               & $2.47^{***}$  & $2.37^{***}$  & $2.45^{***}$  & $2.48^{***}$  & $2.47^{***}$  & $2.47^{***}$  \\
                               &              &               & $(0.16)$      & $(0.16)$      & $(0.16)$      & $(0.17)$      & $(0.17)$      & $(0.17)$      \\
Frac. Hispanic                 &              &               & $3.31^{***}$  & $3.32^{***}$  & $3.00^{***}$  & $3.01^{***}$  & $3.08^{***}$  & $3.08^{***}$  \\
                               &              &               & $(0.18)$      & $(0.18)$      & $(0.19)$      & $(0.19)$      & $(0.19)$      & $(0.19)$      \\
Frac. Asian                    &              &               & $-0.88$       & $-1.38$       & $-1.20$       & $-1.05$       & $-1.29$       & $-1.29$       \\
                               &              &               & $(0.77)$      & $(0.77)$      & $(0.77)$      & $(0.79)$      & $(0.79)$      & $(0.79)$      \\
Frac poverty                   &              &               &               & $2.66^{***}$  & $1.75^{**}$   & $1.77^{**}$   & $1.44^{**}$   & $1.44^{**}$   \\
                               &              &               &               & $(0.54)$      & $(0.55)$      & $(0.55)$      & $(0.55)$      & $(0.55)$      \\
Frac. no health   coverage     &              &               &               &               & $2.73^{***}$  & $2.72^{***}$  & $2.57^{***}$  & $2.57^{***}$  \\
                               &              &               &               &               & $(0.38)$      & $(0.38)$      & $(0.38)$      & $(0.38)$      \\
Frac   owner occupied          &              &               &               &               &               & $0.23$        & $-0.32$       & $-0.32$       \\
                               &              &               &               &               &               & $(0.29)$      & $(0.30)$      & $(0.30)$      \\
Frac mobile   home             &              &               &               &               &               &               & $1.60^{***}$  & $1.60^{***}$  \\
                               &              &               &               &               &               &               & $(0.27)$      & $(0.27)$      \\
\midrule
Num. observations              & $3220$       & $3220$        & $3219$        & $3219$        & $3219$        & $3219$        & $3219$        & $3219$        \\
R$^2$ (full model)             & $0.38$       & $0.42$        & $0.51$        & $0.51$        & $0.52$        & $0.52$        & $0.53$        & $0.53$        \\
R$^2$ (proj model)             & $0.03$       & $0.10$        & $0.23$        & $0.24$        & $0.25$        & $0.25$        & $0.26$        & $0.26$        \\
Adj. R$^2$ (full model)        & $0.37$       & $0.41$        & $0.50$        & $0.50$        & $0.51$        & $0.51$        & $0.52$        & $0.52$        \\
Adj. R$^2$ (proj model)        & $0.01$       & $0.09$        & $0.22$        & $0.23$        & $0.24$        & $0.24$        & $0.25$        & $0.25$        \\
Num. of State Fixed Effects    & $52$         & $52$          & $52$          & $52$          & $52$          & $52$          & $52$          & $52$          \\
\bottomrule
\multicolumn{9}{l}{\scriptsize{$^{***}p<0.001$; $^{**}p<0.01$; $^{*}p<0.05$}}
\end{tabular}

} 

\end{center}

\emph{Sources: County-level confirmed cases as of August 20, 2020,
from the Johns Hopkins Coronavirus Research Center. County population
and other demographic characteristics from the 2018 American Community
Survey. Density is the ratio of ACS population over the area of the
county in squared kilometers using the Census Bureau's boundary shapefile
and the U.S. National Atlas 2163 projected coordinate reference system.}
\end{table}

\clearpage{}

\pagebreak{}

\begin{table}

\caption{Within-City Adaptation: Short-Run Suburbanization in New York, March-July
2020\label{tab:Within-City-Adaptation:-Short-Ru}}
\bigskip{}

\emph{This table uses the ZIP-month Zillow House Value Index (ZHVI)
for the Zip codes of the New York-Newark-Jersey City, NY-NJ-PA Metropolitan
Statistical Area to regress the year-on-year appreciation (in logs)
on the distance to the center (upper panel) and the logarithm of population
density (lower panel). The distance to the center is the kilometer
distance from the centroid of the Zip code tabulation area to the
central business district. Population density computed using the Census
Bureau's 2018 American Community Survey and the 2010 boundaries of
Census Zip code tabulation areas.}

\bigskip{}

\begin{center}

\begin{tabular}{l c c c}
  \toprule
  & \multicolumn{3}{c}{Dependent variable: YoY Price Appreciation} \\
  \cmidrule(lr){2-4}
Time period & 2015--2019 & March-July 2019 & March-July 2020 \\
\midrule
(Intercept)         & $-1.116^{***}$ & $-0.622$     & $2.523^{***}$  \\
                    & $(0.208)$      & $(0.782)$    & $(0.816)$      \\
log(density)        & $0.032$        & $0.179^{**}$ & $-0.327^{***}$ \\
& $(0.021)$      & $(0.091)$    & $(0.094)$      \\
Additional controls & \multicolumn{3}{c}{Year fixed effects} \\
\midrule
R$^2$               & $0.013$        & $0.007$      & $0.022$        \\
Adj. R$^2$          & $0.012$        & $0.005$      & $0.020$        \\
Num. obs.           & $13439$        & $581$        & $541$          \\
\bottomrule
\\
  \toprule
 & \multicolumn{3}{c}{Dependent variable: YoY Price Appreciation} \\
  \cmidrule(lr){2-4}
Time period: & 2015--2019 & March-July 2019 & March-July 2020 \\
\midrule
(Intercept)          & $-0.750^{***}$ & $1.067^{***}$ & $-0.806^{***}$ \\
                     & $(0.102)$      & $(0.204)$     & $(0.211)$      \\
Distance to center (km) & $-0.004^{***}$ & $-0.007$      & $0.023^{***}$  \\
& $(0.001)$      & $(0.005)$     & $(0.006)$      \\
Additional controls & \multicolumn{3}{c}{Year fixed effects} \\
\midrule
R$^2$                & $0.013$        & $0.003$       & $0.024$        \\
Adj. R$^2$           & $0.013$        & $0.001$       & $0.022$        \\
Num. obs.            & $13439$        & $581$         & $541$          \\
\bottomrule
\multicolumn{4}{l}{\scriptsize{$^{***}p<0.01$; $^{**}p<0.05$; $^{*}p<0.1$}}
\end{tabular}

\end{center}
\end{table}

\clearpage{}

\pagebreak{}

\begin{sidewaystable}
\caption{Explaining Metropolitan Growth in the Long-Run: Shocks vs. Fundamentals\label{tab:What-Explains-The-Evolution-of-Cities}}

\emph{This table presents a regression of the change in a metropolitan
area's ranking between 1970 and 2010 on shocks and fundamentals. The
shocks are (i)~billion dollar storms according to NOAA's database
of significant storm events, (ii)~protests with damage to property.
The fundamentals are black-white spatial segregation, education, housing
supply elasticity, industrial composition, industrial diversification. }

\begin{center}
{\scriptsize      

\begin{tabular}{l c c c c c c c c c c c}
\toprule
 & \multicolumn{11}{c}{Dependent variable: Change in metropolitan area population ranking, 1970--2010} \\
\midrule
\\
\textbf{Fundamentals}\\
\\
Black-white dissimilarity                       & $-44.41^{**}$ &                &                &                 &                &               &               &          &          &           & $-47.15^{**}$  \\
                                                & $(15.89)$     &                &                &                 &                &               &               &          &          &           & $(17.35)$      \\
\% College                            &               & $175.58^{***}$ &                &                 &                &               &               &          &          &           & $87.64$        \\
                                                &               & $(52.66)$      &                &                 &                &               &               &          &          &           & $(64.11)$      \\
Housing supply elasticity                                      &               &                & $9.79$         &                 &                &               &               &          &          &           & $9.36^{*}$     \\
&               &                & $(5.33)$       &                 &                &               &               &          &          &           & $(4.58)$       \\

\emph{Industrial composition}\\
\\
\% Mining \& oil                       &               &                &                & $0.06$          &                &               &               &          &          &           & $0.12$         \\
                                                &               &                &                & $(0.25)$        &                &               &               &          &          &           & $(0.27)$       \\
\% Construction                             &               &                &                & $8.45^{***}$    &                &               &               &          &          &           & $5.44^{**}$    \\
                                                &               &                &                & $(1.53)$        &                &               &               &          &          &           & $(2.09)$       \\
\% Manufacturing                            &               &                &                & $0.33$          &                &               &               &          &          &           & $-0.91$        \\
                                                &               &                &                & $(1.19)$        &                &               &               &          &          &           & $(1.80)$       \\
\% Utilities                &               &                &                & $1.22$          &                &               &               &          &          &           & $1.58$         \\
                                                &               &                &                & $(2.45)$        &                &               &               &          &          &           & $(3.19)$       \\
\% Finance/RE                         &               &                &                & $6.42^{**}$     &                &               &               &          &          &           & $3.71$         \\
&               &                &                & $(2.11)$        &                &               &               &          &          &           & $(2.59)$       \\
\% Retail                                   &               &                &                & $1.33$          &                &               &               &          &          &           & $3.30^{*}$     \\
&               &                &                & $(1.02)$        &                &               &               &          &          &           & $(1.41)$       \\
\emph{Industrial specialization} \\
\\
HHI Q2                       &               &                &                &                 & $-5.95$        &               &               &          &          &           & $-4.70$        \\
                                                &               &                &                &                 & $(6.31)$       &               &               &          &          &           & $(6.30)$       \\
HHI Q3                       &               &                &                &                 & $-12.02^{*}$   &               &               &          &          &           & $-6.26$        \\
                                                &               &                &                &                 & $(6.08)$       &               &               &          &          &           & $(7.56)$       \\
HHI Q4                       &               &                &                &                 & $-42.33^{***}$ &               &               &          &          &           & $-34.25^{**}$  \\
&               &                &                &                 & $(8.36)$       &               &               &          &          &           & $(11.82)$      \\
\\
\textbf{Shocks} \\
\\
A riot with damage to property &               &                &                &                 &                & $-2.54$       & $-9.40$       &          &          &           & $-9.60$        \\
                                                &               &                &                &                 &                & $(7.02)$      & $(10.43)$     &          &          &           & $(6.67)$       \\
Nbr of riots with damage to property             &               &                &                &                 &                &               & $4.18$        &          &          &           &                \\
                                                &               &                &                &                 &                &               & $(4.70)$      &          &          &           &                \\
Nbr of b\$ storms                                     &               &                &                &                 &                &               &               & $-0.02$  &          &           &                \\
                                                &               &                &                &                 &                &               &               & $(2.69)$ &          &           &                \\
Property damages                        &               &                &                &                 &                &               &               &          & $-0.02$  &           &                \\
                                                &               &                &                &                 &                &               &               &          & $(0.81)$ &           &                \\
Any b\$ storm                                  &               &                &                &                 &                &               &               &          &          & $-1.89$   & $2.48$        \\
                                            &               &                &                &                 &                &               &               &          &          & $(10.79)$ & $(10.37)$      \\
\midrule
R$^2$                                           & $0.03$        & $0.04$         & $0.07$         & $0.29$          & $0.10$         & $0.09$        & $0.09$        & $0.00$   & $0.00$   & $0.00$    & $0.35$         \\
Num. obs.                                       & $306$         & $306$          & $306$          & $306$           & $306$          & $306$         & $306$         & $306$    & $306$    & $306$     & $306$          \\
\bottomrule
\multicolumn{12}{l}{\scriptsize{$^{***}p<0.001$; $^{**}p<0.01$; $^{*}p<0.05$}}
\end{tabular}

} 
\end{center}  
\vfill

\emph{Sources: NOAA's Storm Events Database, Ethnic Collective Action
in Contemporary Urban United States, 1954-1992 (ICPSR 34341), County
Business Patterns, the National Historical Geographic Information
System tract level Census file of 1970.}
\end{sidewaystable}

\clearpage{}

\pagebreak{}

\begin{table}
\caption{After a Shock: Population Changes in the San Francisco Bay Area After
the 1989 Loma Prieta Earthquake\label{tab:Within-City-Adaptation:-Populati}}

\emph{These six regressions present the regression of decennial log
population change (upper panel) and population rank (lower panel)
on the share of a tract in an earthquake liquefaction area.}

\begin{center}

\begin{tabular}{l c c c}
\toprule
 & \multicolumn{3}{c}{$\Delta$ Census Tract log Population} \\
 \cmidrule(lr){2-4}
 & 1990--1980 & 2000--1990 & 2010--2000 \\
\midrule
(Intercept)         & $0.29^{***}$ & $0.16^{***}$ & $0.10^{***}$ \\
                    & $(0.02)$     & $(0.01)$     & $(0.02)$     \\
\% in liquefaction area & $-0.12^{**}$ & $-0.01$      & $-0.01$      \\
                    & $(0.04)$     & $(0.03)$     & $(0.04)$     \\
\hline
R$^2$               & $0.01$       & $0.01$       & $0.01$       \\
Adj. R$^2$          & $0.01$       & $0.01$      & $0.01$      \\
Num. obs.           & $1,791$       & $1,791$       & $1,791$       \\
\midrule
 & \multicolumn{3}{c}{$\Delta$ Census Tract Population Rank} \\
 \cmidrule(lr){2-4}
 & 1990--1980 & 2000--1990 & 2010--2000 \\
\midrule
(Intercept)         & $7.81$       & $-1.26$   & $1.86$    \\
                    & $(7.64)$     & $(6.61)$  & $(9.20)$  \\
\% in liquefaction area & $-35.90^{*}$ & $5.81$    & $-8.57$   \\
                    & $(18.01)$    & $(15.59)$ & $(21.69)$ \\
\midrule
R$^2$               & $0.01$       & $0.01$    & $0.01$    \\
Adj. R$^2$          & $0.01$       & $0.01$   & $0.01$   \\
Num. obs.           & $1,791$       & $1,791$    & $1,791$    \\
\hline
\multicolumn{4}{l}{\scriptsize{$^{***}p<0.001$; $^{**}p<0.01$; $^{*}p<0.05$}}
\end{tabular}

\end{center}

\emph{Source: California Department of Conservation's regulatory liquefaction
maps. Neighborhood Change Database with 2010 Census Tract Boundaries. }
\end{table}

\end{document}